\documentclass[a4paper,physrev,twocolumn]{revtex4-2}
\usepackage{amssymb}
\usepackage{amsmath}
\usepackage{bm}
\usepackage{braket}
\usepackage{graphicx}
\graphicspath{ {./figs/} }
\usepackage{color}
\usepackage{url}
\usepackage{siunitx}
\usepackage{hyperref}
\hypersetup{
  colorlinks=true,
  linkcolor=blue,
  filecolor=blue,      
  citecolor=blue,
  urlcolor=blue,
}
\newcommand{\I}{\mathrm{i}}
\newcommand{\E}{\mathrm{e}}
\newcommand{\D}{\mathrm{d}}
\newcommand{\BZ}{\mathcal{B}}

\DeclareMathOperator{\Tr}{tr}
\DeclareMathOperator{\CZ}{CZ}

\begin{document}

\title{Entanglement dynamics and phase transitions of the Floquet cluster spin chain}
\author{Alberto D.\ Verga}\email{alberto.verga@univ-amu.fr}
\affiliation{CPT, Aix-Marseille Université, Marseille, France}

\date{\today}
\begin{abstract}
Cluster states were introduced in the context of measurement based quantum computing. In one dimension, the cluster Hamiltonian possesses topologically protected states. We investigate the Floquet dynamics of the cluster spin chain in an external field, interacting with a particle. We explore the entanglement properties of the topological and magnetic phases, first in the integrable spin lattice case, and then in the interacting quantum walk case. We find, in addition to thermalization, dynamical phase transitions separating low- and high-entanglement nonthermal states, reminiscent of the ones present in the integrable case, but differing in their magnetic properties. The nonergodic phases are characterized by the emergence of magnetic order, persistent at long times.
\end{abstract}
\maketitle

\section{Introduction}
\label{S:intro}

One of the main trends in quantum information is the search for ``computational phases'' of matter \cite{Doherty-2009,Stephen-2019}. Indeed, since the formulation of models of fault-tolerant \cite{Shor-1996,Preskill-1998}, robust-against-error \cite{Gottesman-1997}, quantum computation using logical qubits encoded in the degenerate ground state of gapped Hamiltonians  and their topological excitations \cite{Kitaev-2003fk}, or using persistent, highly entangled quantum states \cite{Briegel-2001fk}, different models of quantum computation have been introduced. The main idea has been to use the properties of topological phases of matter, protected by some kind of symmetry to implement the qubit logical operations, as in the topological model \cite{Freedman-2003}, or the measurement-based model \cite{Raussendorf-2001uq} of quantum computation.

One obstacle in the way of self-correcting fault-tolerant quantum computation is the need for robust topologically ordered phases supporting highly entangled states as a universal quantum resource \cite{Dennis-2002,Brown-2016,Roberts-2019,Wildeboer-2022}. In fact, closed quantum systems with short range, local interactions, conserving only total energy, tend to thermalize: The expected value of the observables is essentially given by their microcanonical value as derived from the eigenvector thermalization \cite{Alessio-2016fj}.

As a result of thermalization, the states, although highly entangled, cannot be a useful resource for quantum computation \cite{Gross-2009uq}. If, in addition, the system is periodically driven by an applied field, breaking then the energy conservation, it should evolve towards an infinite temperature ergodic state \cite{DAlessio-2014,Lazarides-2014}. In systems without extrinsic disorder, protection against thermalization can also be reached by ergodicity breaking, in analogy with classical glasses, due to the existence of additional conservation laws \cite{Chamon-2005,Sala-2020,Scherg-2021a}, or by the emergence of a decoupled subspace of nonthermal states \cite{Shiraishi-2017,Papic-2022}.

In Floquet systems, in which we are interested, exceptions to relaxation towards a thermal state are integrable systems whose large number of local constants of motion prevent the emergence of an ergodic phase \cite{Gopalakrishnan-2018,Friedman-2019}, and systems with dynamical constraints, as in arrays of Rydberg atoms \cite{Bernien-2017} supporting many-body scar states \cite{Iadecola-2020,Mizuta-2020,Sugiura-2021}. Another possibility for suppression of unbounded energy absorption in strongly driven systems, is the emergence of a robust (albeit not exact) conserved quantity leading to dynamical freezing \cite{Das-2010}. For a recent review see Ref.~\cite{Haldar-2022}.

The search for nonergodic phases in Floquet systems is motivated by the possibility of using them to engineer effective Hamiltonians in order to describe, for instance, topological materials \cite{Oka-2019} or, from a more fundamental perspective, to investigate nonequilibrium phases of matter \cite{Harper-2020,Yates-2022,Wybo-2021}. One interesting possibility is to create, using Floquet dynamics, nonthermal states possessing useful entanglement properties, similar to the ones found in symmetry-protected topological phases of gapped Hamiltonians \cite{Zeng-2019}, having the potential to be a universal resource for quantum computation, such as for instance the cluster state \cite{Nielsen-2006fv,Raussendorf-2012xr}.

Breaking of ergodicity in Floquet systems was recently demonstrated in the case of a periodically perturbed ergodic Ising spin chain \cite{Haldar-2018,Haldar-2021}, and in the case of a quantum cellular automaton \cite{Sellapillay-2022b}. Although the two models are unrelated, the mechanism of ergodicity breaking have in common the emergence of approximate integrals of motion \cite{Das-2010}: the magnetization in the first case \cite{Haldar-2018} and the conservation of the quasiparticle number in the second case \cite{Sellapillay-2022b}.

For generic Ising chains, the emergence of an approximate conservation law, absent in the undriven system, is related to the presence of resonances. These resonances are a many-body generalization of the single-particle resonances that, by interference at particular driven frequencies, coherently destroy amplitude tunneling \cite{Grossmann-1991}, leading to energy eigenstate localization. At strong driving, these resonances effectively create constraints (similar to the ones in scar models), eventually leading to the freezing of the system dynamics and the breaking of the system's ergodicity \cite{Haldar-2021}. In contrast to the Ising chain, ergodicity breaking in the automaton, is due to the persistence of weakly interacting quasiparticles in the chaotic regime, even in the absence of energy eigenstate localization, leading to the fragmentation of the Hilbert space according to their content in quasiparticles \cite{Sellapillay-2022b}. 

Our aim in this paper is to investigate the fate of a topological phase associated with the ground state of an error-correcting Hamiltonian, such as the cluster phase \cite{Doherty-2009,Son-2011}, first when externally driven by a periodic field, and then when embedded in a larger Hilbert space by the introduction of an interaction with a quantum walker \cite{Kitagawa-2012fk}. More specifically, we generalize the Floquet cluster model, which is integrable, by introducing an exchange coupling between the chain spins and the walker spin (the coin internal degree of freedom). We chose a coupling such that it preserves the original symmetries, but destroys the local integrals of motion of the cluster chain. The effect of the walker is to mediate, through its ballistic motion, the interaction between spatially separated spins, effectively adding a nonlocal interaction \cite{Verga-2019,Verga-2019b}.

We focus on a one dimensional spin chain whose Floquet dynamics can be solved analytically (Sec.~\ref{S:model}), and demonstrate the existence of a topological phase issued from the original static phase; in addition, computing the Loschmidt rate \cite{Heyl-2015} we show that the integrable model undergoes a dynamical phase transition \cite{Heyl-2013,Heyl-2018} that can be characterized by the concomitant change in the entanglement level \cite{DeNicola-2021,Jafari-2021}.

The extension of the model to an interacting quantum walk \cite{Verga-2019,Sellapillay-2022} allows us to determine the persistence of the two phases present in the integrable case, the cluster phase and the topologically trivial paramagnetic phase (Sec.~\ref{S:QW}). In fact we find that, even if the dynamical phase transition is always present in a range of parameters, new nonergodic phases appear characterized by a finite magnetization and global entanglement. We discuss the mechanism of ergodicity breaking in terms of the effective magnetic interaction mediated by the particle between fixed spins. We show in particular, that a phase transition between low- and high-entanglement phases, is possible even for the case corresponding to the trivial phase in the original cluster chain.

\section{Floquet cluster model}
\label{S:model}

We consider a system of \(L\) spins in a one dimensional lattice. The system's Hamiltonian is \cite{Raussendorf-2005}
\begin{equation}
\label{e:Hcluster}
H_C = -\frac{J}{2}\sum_{x=0}^{L-1} Z_{x-1} X_x Z_{x+1}  ,
\end{equation}
where \(J\) is the coupling constant of the ``cluster'' interaction; we denote \(\bm \sigma_x = (X_x,Y_x,Z_x)\) the vector of Pauli matrices at each site \(x = 0, \ldots, L-1\) (we take the lattice constant as the unit of length). We assume periodic boundary conditions, to ensure translation invariance. In addition, an external field \(B\) applies in the \(x\) direction:
\begin{equation}
\label{e:Hfield}
H_B = -\frac{B}{2} \sum_{x=0}^{L-1} X_x .
\end{equation}
The spin Hilbert space is spanned by the basis states \(\ket{s} = \ket{s_0 \cdots s_{L-1}}\), \(s_x = \{0,1\}\), labeled by the integer \(s = 0, \ldots, 2^L-1\). The ground state of \(H_C\), the cluster state, is the eigenvector with eigenvalue \(1\) common to every term in \(H_C\); it can be written in terms of the controlled \(Z\) operator \(\CZ = \mathrm{diag}(1,1,1,-1)\):
\begin{equation}
\label{e:Cstate}
\ket{C} = \prod_x \CZ_{x,x+1} \ket{+}^L
\end{equation}
where \(\ket{+} = (\ket{0}+\ket{1})/2^{1/2}\) (we use the tensor product notation ``\(\otimes\)'' only in ambiguous cases, or to separate Hilbert subspaces). The \(H_C\) Hamiltonian is in fact a sum of stabilizer operators \cite{Gottesman-1997}, whose terms commute with each other and commute with the Hamiltonian, hence providing an extensive number of integrals of motion.

The cluster state \eqref{e:Cstate} is a highly entangled quantum state that can also be defined over an arbitrary graph of qubits, used as an information resource for measurement based quantum computing \cite{Briegel-2001fk,Raussendorf-2001uq}. Highly entangled means that its Schmidt dimension, the minimal number of parameters needed for its specification, grows exponentially with the number of qubits. It also has the property of maximal connectedness: Any pair of qubits can be projected into a Bell state by appropriately measuring the intermediate qubits joining them along a path on the graph \cite{Hein-2006eu}. A consequence of the exponential Schmidt rank and the inherent entanglement nonlocality, the cluster state is a genuine quantum resource that goes beyond the possibilities of any classical resource: A computation based on the cluster state cannot, in principle, be efficiently simulated by a classical algorithm \cite{Vidal-2003}.

The dynamics of the system is governed by the Floquet operator
\begin{equation}
\label{e:Flo}
\begin{split}
F(J,B) &= \E^{-\I H_C} \E^{-\I H_B} \\
&= \prod_{x=0}^{L-1} \E^{\I (J/2) Z_{x-1} X_x Z_{x+1}} \prod_{x=0}^{L-1} \E^{\I (B/2) X_x}\,
\end{split}
\end{equation}
such that the state of the system \(\ket{\psi(t)}\) at time \(t\) changes in one time step (our unit of time, with \(\hbar=1\)) according to
\begin{equation}
\label{e:Fonet}
\ket{\psi(t+1)} = F(J,B) \ket{\psi(t)} .
\end{equation}
The parity operators
\begin{equation}
\label{e:parity}
P_o = \prod_{x\in o} X_x, \quad P_e = \prod_{x\in e} X_x, \quad P = P_o P_e,
\end{equation}
where \(o,e\) stand for odd and even sites, commute with \(F\):
\begin{equation}
\label{e:PePo}
[P_e,F] = [P_o,F] = [P,F].
\end{equation}
Therefore, \(H_C\) is symmetric under the \(\mathbb{Z}_2 \times \mathbb{Z}_2\) group generated by \(P_{e,o}\), and its ground state belongs to a topological phase protected by symmetry \cite{Son-2011,Verresen-2017}. This suggests the question about the fate of this topological phase in the Floquet case.

Both the Hamiltonian \(H_C+H_B\) and the Floquet \(F\) models possess the same \(\mathbb{Z}_2 \times \mathbb{Z}_2\) symmetry; however \(F\) do not commute with \(H_C+H_B\) breaking the energy conservation, which leads to essentially different dynamical properties. The Hamiltonian model, often called the transverse field cluster model, was extensively analyzed, starting with the calculation by Suzuki \cite{Suzuki-1971} who used the Jordan-Wigner transformation to find the eigenspectrum of \(H_C+H_B\), and followed by the description of the phase transition between the \(J \rightarrow 0\) paramagnetic phase and the \(B \rightarrow 0\) topological phase \cite{Pachos-2004,Doherty-2009}, and the discussion of the phase diagram when magnetic interactions are added \cite{Skrovseth-2009,Smacchia-2011,Son-2011,Montes-2012}.

To find the spectrum of \(F\) we follow the usual method \cite{Suzuki-1971,Lakshminarayan-2005} and introduce the Jordan-Wigner transformation \cite{Jordan-1928,Mbeng-2020} of the spin operators in terms of the fermion operators,
\begin{equation}
\label{e:ff}
\{f_x,f_y\} = \{f_x^\dagger,f_y^\dagger\}, \quad \{f_x,f_y^\dagger\} = \delta_{xy}.
\end{equation}
It is defined by
\begin{equation}
\begin{split}
X_x &= 1 - 2 f^\dagger_x f_x, \\
Y_x &= \I K_x (f^\dagger_x - f_x), \\
Z_x &= -K_x (f^\dagger_x + f_x),
\end{split}
\label{e:JW}
\end{equation}
where
\begin{equation}
\label{e:Kff}
K_x = \prod_{y=0}^{x-1} \E^{\I \pi f^\dagger_y f_y} = \prod_{y=0}^{x-1}( 1 - 2 f^\dagger_y f_y ).
\end{equation}
Once introduced into \eqref{e:Hcluster}, the Jordan-Wigner transformation leads to the bilinear fermion Hamiltonian
\begin{align}
\label{e:HCBf}
H_C &= \frac{J}{2} \sum_{x=0}^{L-1} (f^\dagger_{x-1} - f_{x-1})(f^\dagger_{x+1} + f_{x+1}), \\
H_B &= \frac{B}{2} \sum_{x=0}^{L-1} (2f^\dagger_x f_x - 1).
\end{align}
The translation invariance allows the use of the Fourier transform
\begin{equation}
\label{e:Fourier}
f_x = \frac{\E^{\I \pi/4}}{\sqrt{L}} \sum_k \E^{\I k x} f_k,
\end{equation}
where the set of wavenumbers in the Brillouin zone \(k \in (-\pi, \pi]\), is divided into ``even'' and ``odd'' sectors,
\begin{equation}
\begin{split}
e &= \{ k = \pm \frac{\pi n}{L} \mid n = 1,3,\ldots, L-1 \} \\ 
&= e_+ \cup e_-, \\
o &= \{ k = \frac{\pi n}{L} \mid n = 0,\pm 2,\ldots, \pm (L-2), L \} \\
&= \{0\} \cup o_+ \cup o_- \cup \{\pi\},
\end{split}
\label{e:keo}
\end{equation}
for an even and odd number of fermions, respectively (taking \(L\) to be even). Indeed, the sign of the parity operator \(P\) depends on the total number of fermions \(N_F\),
\begin{equation}
\label{e:Pff}
P = (-1)^{N_F}, \quad N_F = \sum_{x=0}^{L-1} f_x^\dagger f_x,
\end{equation}
for even \(N_F\), we impose anti-periodic boundary conditions \(f_{L+1} = -f_1\), and we impose periodic boundary conditions \(f_{L+1}=f_1\), for \(N_F\) odd. The Fourier transformed cluster Hamiltonian is, in the even sector,
\begin{multline}
\label{e:HCk}
H_C = J \sum_{k \in e_+} \left[ \cos 2k ( f^\dagger_k f_k - f_{-k} f^\dagger_{-k} ) \right. \\ \left. + \sin 2k ( f^\dagger_{k} f^\dagger_{-k} + f_{-k} f_k ) \right] ,
\end{multline}
and the field Hamiltonian is,
\begin{equation}
\label{e:HBk}
H_B = B \sum_{k \in e_+} ( f^\dagger_k f_k - f_{-k} f^\dagger_{-k} ) .
\end{equation}
Similar expressions hold for the odd sector.

The unitary map \(F\) is, in the even sector
\begin{equation}
\label{e:Fke}
F^{(e)} = \prod_{k \in e_+} V_k \prod_{k \in e_+} W_k ,
\end{equation}
where
\begin{equation}
\label{e:VkWk}
V_k = \E^{\I J C_k^\dagger (\cos 2k Z + \sin 2k X) C_k}, \quad
W_k = \E^{\I B C_k^\dagger Z C_k} ,
\end{equation}
and
\begin{equation}
\label{e:Ck}
C_k = \begin{pmatrix} f_k \\ f^\dagger_{-k} \end{pmatrix}, \quad
C^\dagger_k = \begin{pmatrix} f^\dagger_k & f_{-k} \end{pmatrix}.
\end{equation}
In the odd sector we add the terms \(k=0, \pi\):
\begin{equation}
\label{e:Fko}
F^{(o)} = F_0 \prod_{k \in o_+} V_k \prod_{k \in o_+} W_k F_\pi ,
\end{equation}
where
\begin{equation}
\label{e:F0pi}
F_0 = \E^{\I(J+B)(2 f^\dagger_0 f_0 - 1)}, \quad
F_\pi = \E^{\I(J+B)(2 f^\dagger_\pi f_\pi - 1)} .
\end{equation}
Combining the two rotations we obtain the Floquet operator in terms of an effective Hamiltonian \(H_F\):
\begin{equation}
\label{e:FHeff}
F^{(e)} = \E^{-\I H_F}, \; F^{(o)} = F_0 \E^{-\I H_F} F_\pi ,
\end{equation}
where,
\begin{equation}
\label{e:HF}
H_F = -\sum_k C^\dagger h_k C_k 
\end{equation}
and
\begin{equation}
\label{e:hk}
h_k = \epsilon_k \bm n_k \cdot \bm \sigma = \bm d_k \cdot \bm \sigma ,
\end{equation}
with \(\bm d_k = \epsilon_k \bm n_k\) and
\begin{equation}
\label{e:ek}
\cos(\epsilon_k) = \cos J \cos B - \sin J \sin B \cos 2k;
\end{equation}
\begin{equation}
\label{e:nk}
\bm n_k = \frac{1}{\big| \sin \epsilon_k \big|} \begin{pmatrix}
\sin J \cos B \sin 2k \\
\sin J \sin B \sin 2k \\
\sin J \cos B \cos 2k + \cos J \sin B \end{pmatrix}
\end{equation}
is a unit vector of \((n_x, n_y, n_z)\) components. Note that similar effective Hamiltonians appear in the Floquet transverse field Ising model \cite{Lakshminarayan-2005}, and topological quantum walks \cite{Kitagawa-2012fk}.

\begin{figure}
  \centering
  \includegraphics[width=0.85\columnwidth]{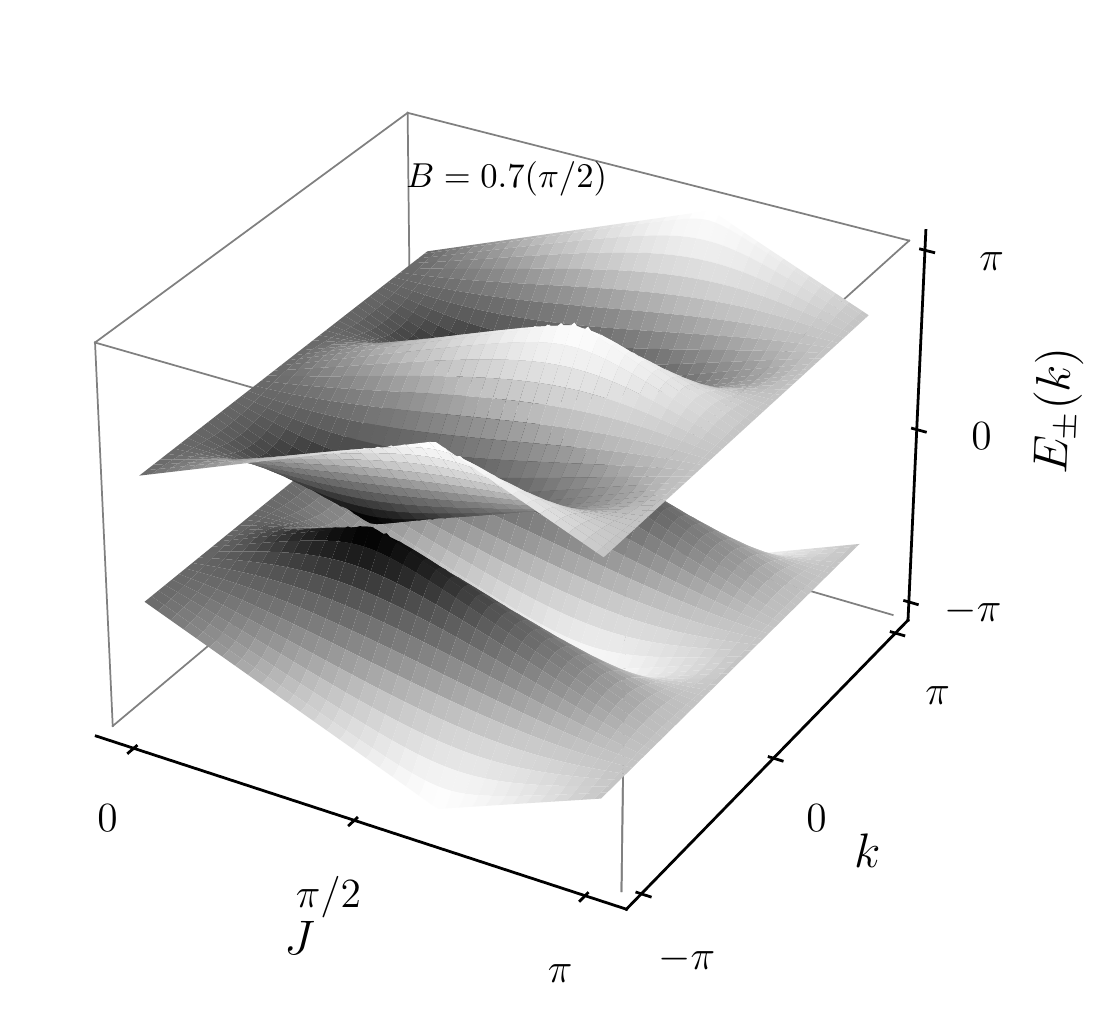}
  \caption{Family of quasienergy bands $\epsilon_k$ of the effective Hamiltonian, as a function of the quasimomentum $k$ and $J$, for $B=0.7\pi/2$, fixed. The gap between positive and negative quasienergy bands closes on the line $J=B$, at $\epsilon_k=0$ when $k = \pm\pi/2$, and at $\epsilon_k=\pi$ when $k=0, \pi$.
  \label{f:ek}}
\end{figure}

\subsection{Winding number}

The dispersion relation \(\epsilon_k\) \eqref{e:ek}, is represented in Fig.~\ref{f:ek} as a function of \(J\) for fixed \(B\); two Dirac points appear in the Brillouin zone \(k\in (-\pi,\pi]\) when \(J=B\) for \(\epsilon_k=0, \pi\). We demonstrate now that the change in the band structure through the line in parameter space \(J=B\) separates two topological distinct phases, related to a chiral symmetry of the effective Hamiltonian.

\begin{figure}[tb]
  \centering
  \includegraphics[width=0.5\columnwidth]{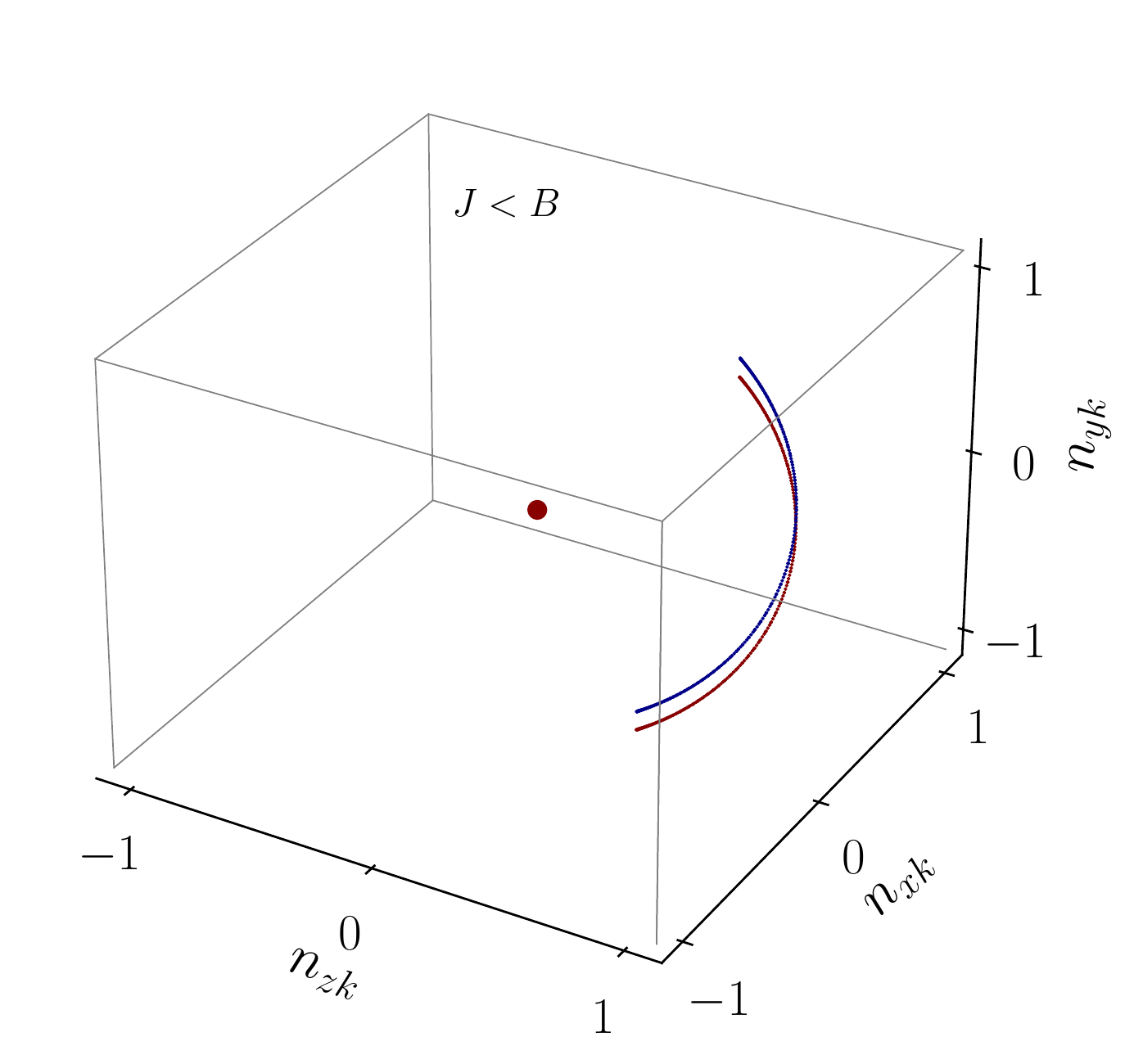}%
  \includegraphics[width=0.5\columnwidth]{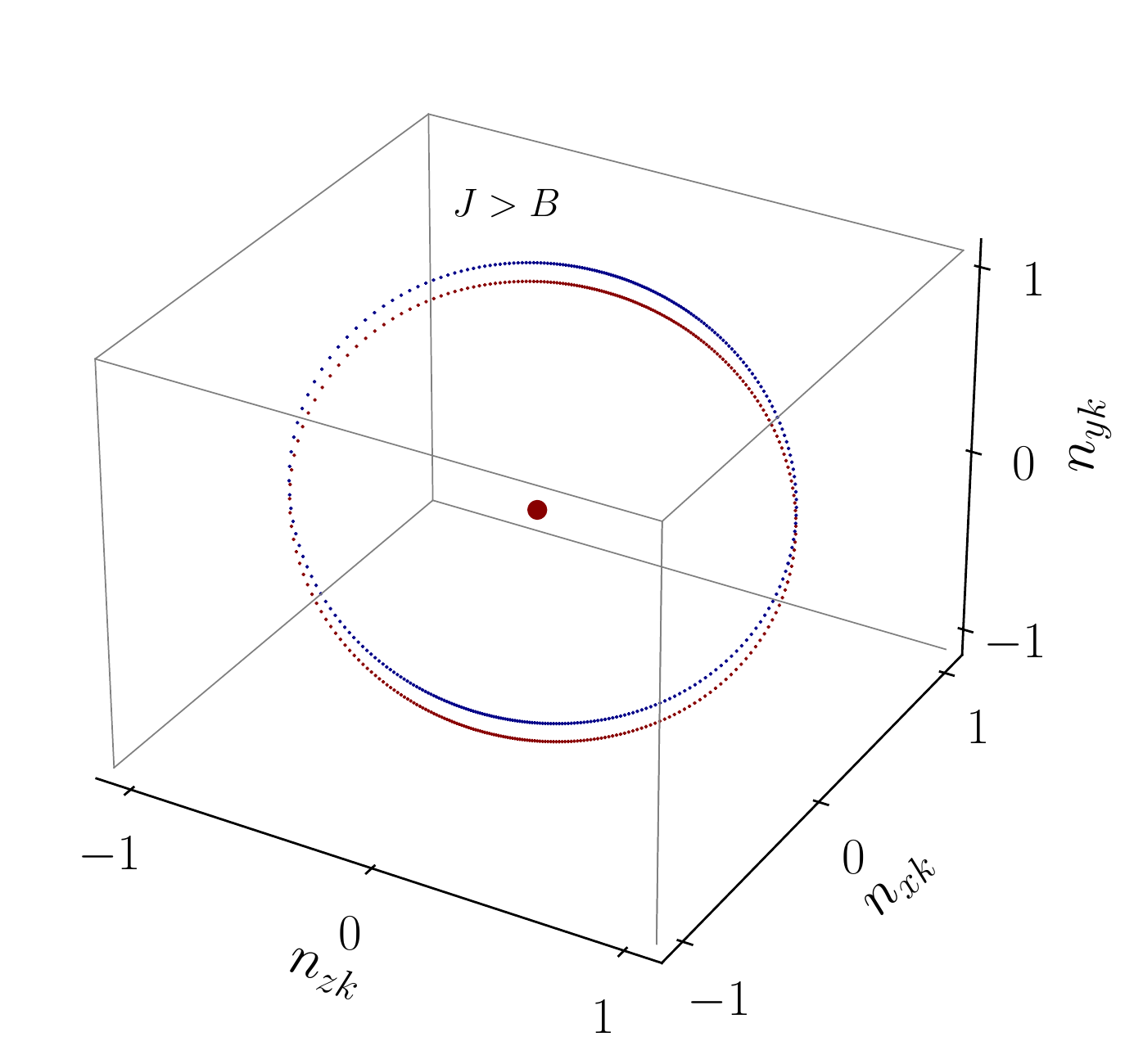}
  \caption{The rotation axis vector of the effective Hamiltonian accumulates a zero phase for $J<B$ and a $4\pi$ phase for $J>B$, when $k$ spans the Brillouin zone. (The two circles are shifted for clarity; the dot marks the origin of coordinates.) Parameters: $J>B$ case, $J=0.9\pi/2$, $B=0.7\pi/2$, and $J<B$ case, $J=0.5\pi/2$, $B=0.7\pi/2$.
  \label{f:nk}}
\end{figure}

We show in Fig.~\ref{f:nk} the locus of \(\bm n_k\) for two values of \((J,B)\). When the spin coupling constant is small with respect to the applied field \(J<B\), the vector describes an arc, and in the opposite case it describes a complete circle; both run twice around the center. 

We note that the vector \(\bm A\)
\begin{equation}
\label{e:vecA}
\bm A = \bm A(B) = \begin{pmatrix}\sin(B) \\ -\cos(B) \\ 0 \end{pmatrix}
\end{equation}
is perpendicular to \(\bm n_{\bm k}\), independently of \(k\). The existence of such vector is related to the symmetry
\begin{equation}
\label{e:AhA}
\E^{\I \pi \bm A \cdot \bm \sigma/2} h \E^{-\I \pi \bm A \cdot \bm \sigma/2} = -h,
\end{equation}
indicative of a chiral invariance of the effective Hamiltonian. Indeed, applying the rotation
\begin{equation}
\label{e:rot}
\E^{-\I \pi X/4}\E^{\I B Y/2} h \E^{-\I B Y/2} \E^{\I \pi X/4} = h_c ,
\end{equation}
we can off-diagonalize the effective Hamiltonian, explicitly exhibiting its chiral symmetry:
\begin{equation}
\label{e:hc}
h_c = \begin{pmatrix}0 & g(k) \\ \bar{g}(k) & 0 \end{pmatrix} = \bm d_c(k) \cdot \bm \sigma  ,
\end{equation}
where \(g(k) = d_{cx}(k) - \I d_{cy}(k)\), and where
\begin{equation}
\label{e:dc}
\bm d_c = \frac{\epsilon_k}{\sin \epsilon_k}\begin{pmatrix}\sin J \sin 2k \\ \sin J \cos B \cos 2k + \cos J \sin B \\ 0\end{pmatrix}
\end{equation}
lies on the $(x,y)$ plane. Therefore, the winding number can be easily computed using the standard formula \cite{Asboth-2016zr}
\begin{equation}
\label{e:nu}
\nu = \frac{1}{2\pi \I} \int_{-\pi}^\pi \D k \, \frac{\D}{\D k} \ln g(k)  .
\end{equation}
(Note that the wavenumber can be extended to the whole Brillouin zone \(k \in (-\pi,\pi]\), since \(\epsilon_k\) is even.)

\begin{figure}[tb]
  \centering
  \includegraphics[width=0.55\columnwidth]{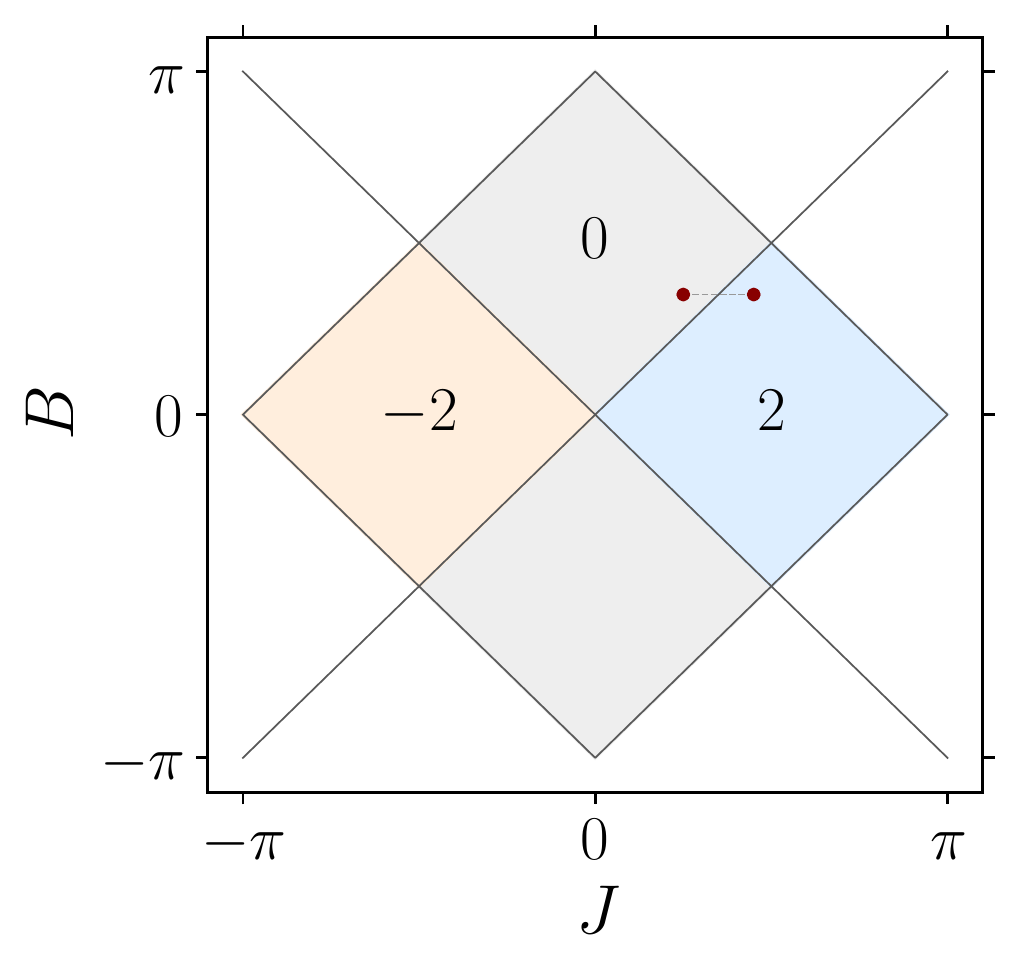}
  \caption{Winding number as a function of the coupling $J$ and field $B$. The two dots correspond to the trivial and nontrivial phases of Fig.~\protect\ref{f:nk}.
  \label{f:JBtop}}
\end{figure}

Noting that the coordinates of \(\bm d_c\) define the parametric equations of an ellipse centered at \((0, \cos J \sin B)\), we obtain
\begin{equation}
\label{e:nu2}
\nu = \begin{cases} 2\operatorname{sgn}(\sin J ) & \text{if } \tan J > \tan B \\ 0 & \text{if } \tan J < \tan B\end{cases}  ,
\end{equation}
the critical points locate at the lines \(J = B \mod \pi/2\), splitting the \((J,B)\) plane into sectors with winding number \(\nu = 0, \pm2\) (Fig.~\ref{f:JBtop}).

We remark that the use of \eqref{e:nu} to characterize the topological phases is based on the symmetries of the effective Hamiltonian, in particular the winding number is invariant under the unitary transformation \eqref{e:rot} leading to the manifestly chiral form \(h_c\) \eqref{e:hc}. However it is possible to generalize \(\nu\) from the symmetries of the Floquet operator instead of those related to the effective Hamiltonian \cite{Roy-2017a}. More specifically, one may take into account the fact that \(\epsilon_k\) is a quasienergy and define a pair of winding numbers associated with the closing of band gaps at \(\epsilon_k(J,B) = 0, \pi\) \cite{Asboth-2013yq,Zhou-2018}. The result \eqref{e:nu2} shows that the Floquet generalization of the cluster model does not destroy the cluster nontrivial topology \cite{Ohta-2016,Verresen-2017}.

\subsection{Eigenstates}

The diagonalization of the effective Hamiltonian can be obtained by a unitary Bogoliubov transformation
\begin{equation}
\label{e:Bogo}
R_k = \frac{1}{\sqrt{2}\sqrt{1-n_z(k)}}  \begin{pmatrix} 1-n_z(k) & n_-(k) \\ -n_+(k) & 1-n_z(k) \end{pmatrix} ,
\end{equation}
where \(n_\pm = n_x \pm \I n_y\) and the columns of $R_k$ are the eigenvectors of \(h_k\) corresponding to the eigenvalues \(\mp \epsilon_k\), respectively. This transformation maps the Fermi operators \(C_k\) into the new operators
\begin{equation}
\label{e:CtoA}
A_k = \begin{pmatrix} a_k \\ a^\dagger_{-k}\end{pmatrix} = R_k^\dagger\begin{pmatrix} f_k \\ f^\dagger_{-k}\end{pmatrix}
\end{equation}
preserving the commutation relations. In the new basis the effective Hamiltonian reads,
\begin{equation}
\label{e:HA}
H_F = \sum_{k>0} A^\dagger_k \begin{pmatrix} \epsilon_k & 0  \\ 0 & -\epsilon_k \end{pmatrix} A_k .
\end{equation}
The corresponding evolution operator factorizes as
\begin{equation}
\label{e:Fk}
F = \left\{
\begin{aligned}
&\prod_{k \in e_+} F_k  , \quad \text{even}\\
&F_0 \left(\prod_{k \in o_+} F_k \right) F_\pi  , \quad \text{odd}
\end{aligned} \right.
\end{equation}
where,
\begin{equation}
\label{e:FkVW}
F_k = \E^{\I C^\dagger_k h_k C_k} .
\end{equation}
Its eigenstates span a four dimensional space for each \(k\), corresponding to the four eigenvalues of the fermion number operators \(f^\dagger_k f_k\) and \(f^\dagger_{-k} f_{-k}\). The basis of this four dimensional space is
\begin{equation}
\label{e:eigenv}
\big\{ \ket{0}, f^\dagger_k \ket{0}, f^\dagger_{-k} \ket{0}, f^\dagger_{-k} f^\dagger_k \ket{0} \big\}
\end{equation}
where \(f_k\ket{0}=f_{-k}\ket{0}=0\).  The vacuum state \(\ket{0}\) corresponds to the completely polarized state \(\ket{+}^L\) in the original spin configuration basis [cf.\ \eqref{e:Cstate}]. In this basis, the eigenstates of \(F_k\) are,
\begin{gather}
F_k \ket{\pm k} = \ket{\pm k}, \nonumber \\
F_k \ket{\pm kk} = \E^{\pm \I \epsilon_k}\ket{\pm kk}, 
\label{e:eigenvF}
\end{gather}
where
\begin{gather}
\ket{\pm k} = f^\dagger_{\pm k} \ket{0}, \nonumber \\
\ket{+ kk} = \frac{1 - n_z(k) + n_-(k) f^\dagger_{-k} f^\dagger_k}{\sqrt{2(1-n_z(k))}}\ket{0}, \nonumber \\
\ket{- kk} = \frac{n_+(k) - (1 - n_z(k)) f^\dagger_{-k} f^\dagger_k}{\sqrt{2(1-n_z(k))}}\ket{0} ,
\label{e:kk}
\end{gather}
(see Appendix~\ref{S:calc}). The \(k = \{0, \pi\}\) subspace is spanned by the eigenvectors
\begin{equation}
\label{e:eigenv0}
\big\{ \ket{0}, f^\dagger_0 \ket{0}, f^\dagger_{\pi} \ket{0}, f^\dagger_0 f^\dagger_{\pi} \ket{0} \big\} ,
\end{equation}
which also are the eigenvectors of \(F_0F_{\pi}\) with eigenvalues:
\begin{equation}
\label{e:eigenval0}
\big\{ \E^{-2\I(J+B)}, \E^{-\I(J+B)}, \E^{-\I(J+B)}, 1 \big\} ,
\end{equation}
respectively. Therefore, these terms contribute with a constant phase and do not play any dynamical role.

\subsection{Time evolution}

The discrete time evolution of an arbitrary state \(\ket{\psi(t)} = F(t) \ket{\psi(0)}\), for integer \(t\), is governed by,
\begin{equation}
\label{e:Ft}
F(t) = \prod_k F_k(t) = \prod_k \E^{\I t \epsilon_k C_k^\dagger \bm n_k \cdot \bm \sigma C_k} ,
\end{equation}
where the product is over the relevant set of wavenumbers [cf.\ \eqref{e:keo} and \eqref{e:Fk}].
For the vacuum state \(\ket{0}\), a simple calculation using the eigenstates \eqref{e:kk}, gives (see Appendix~\ref{S:calc})
\begin{multline}
\label{e:0t}
\ket{t} = \prod_k \big[ \cos(\epsilon_k t) + \I n_z(k) \sin(\epsilon_k t)  \\
+ \I n_-(k) \sin(\epsilon_k t) f_{-k}^\dagger f_k^\dagger \big] \ket{0} .
\end{multline}
We are interested in the time evolution of the entanglement, when the system is initially in the product state \(\ket{0}\) (or equivalently \(\ket{+}^L\)). A computable measure of the global entanglement can be defined in terms of the spin purity \(\Tr \rho_x^2\), where \(\rho_x\) is the reduced density matrix of the spin located at \(x\), assuming that the system is in an arbitrary pure state \(\ket{\psi(t)}\) \cite{Brennen-2003a}:
\begin{equation}
\label{e:Qpure}
\mathcal{Q}(t) = 2 - \frac{2}{L} \sum_{x=0}^{L-1} \Tr \rho^2_x(t).
\end{equation}
This formula can easily be written as
\begin{equation}
\label{e:Q}
\mathcal{Q}(t) = 1 - \frac{1}{L} \sum_x \braket{\psi(t)|\bm \sigma_x |\psi(t)}^2,
\end{equation}
where, from the expression of the density matrix,
\begin{equation}
\label{e:rhox}
\rho_x = \frac{1 + \braket{\bm \sigma_x} \cdot \bm \sigma}{2} ,
\end{equation}
we derived the purity in terms of the expected value of the spin at site \(x\), \(\braket{\bm \sigma_x}\) \cite{Lakshminarayan-2005}. The $\mathcal{Q}$ measure, first introduced in Ref.~\cite{Meyer-2002}, quantifies the multipartite entanglement of a given state as an average over the bipartite entanglement of each spin with the rest of the system \cite{Brennen-2003a}. \(\mathcal{Q}\) vanishes only if the global state is a product state, and is maximum for a globally entangled state. It distinguishes localized and chaotic random states \cite{Giraud-2007} and is useful to characterize the quantum phase transition of the XY model \cite{Radgohar-2018}. The time evolution of \(\mathcal{Q}\) in the kicked Ising model, extensively investigated by Lakshminarayan and Subrahmanyam in Ref.~\cite{Lakshminarayan-2005}, shows recurrences to an initial unentangled state and smooth variations for zero field, but at nonzero transverse field, its dynamics becomes very complex due to the large energy spectrum (time scales) of the Floquet operator (even if the system is completely integrable).

\begin{figure}
  \centering
  \includegraphics[width=0.85\columnwidth]{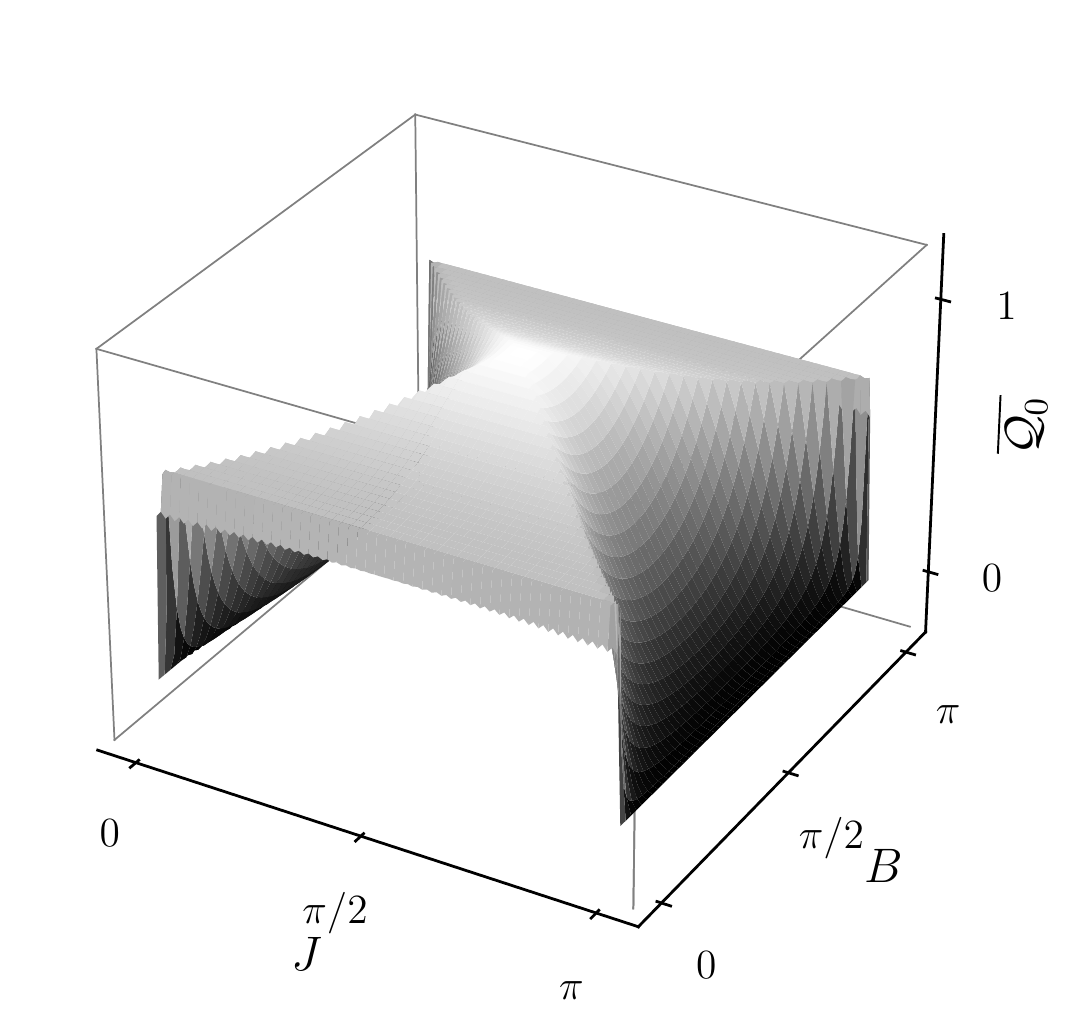}
  \caption{Mean global entanglement \(\mathcal{Q}_0\), averaged over the last 50 time steps (over 200), as a function of \(J\) and \(B\). Computed from \protect\eqref{e:Q0t} with \(1000\) values of \(k\), \(L=1000\).
  \label{f:Q0}}
\end{figure}

Equations \eqref{e:JW} and \eqref{e:0t} imply that the expected values in the evolved vacuum state of \(\braket{Y_x}(t)\) and \(\braket{Z_x}(t)\), vanish, therefore
\begin{equation}
\label{e:Q0}
\begin{split}
\mathcal{Q}_0(t) &= 1 - \frac{1}{L} \sum_x \braket{t|X_x|t}^2 \\ 
&= \frac{4}{L}\sum_x \braket{t|f_x^\dagger f_x|t} \big( 1 - \braket{t|f_x^\dagger f_x|t} \big)  ,
\end{split}
\end{equation}
which gives (see Appendix~\ref{S:calc}),
\begin{equation}
\label{e:Q0t}
\mathcal{Q}_0(t) = \frac{4}{L} \sum_{k\in \BZ} N_k \left[ 1 - 
\frac{1}{L} \sum_{k\in \BZ} N_k \right]  ,
\end{equation}
where 
\begin{equation}
\label{e:Nk}
N_k = \braket{t |f^\dagger_k f_k| t} = (1-n_z^2) \sin^2(\epsilon_k t)  . 
\end{equation}
In the case \(J=B=\pi/2\), \(n_z(k)=0\) and \(\epsilon_k = 2k\) (Dirac dispersion) the global entanglement for the initial \(\ket{0}\) state, reduces to
\begin{equation}
\label{e:Q_0pi}
\mathcal{Q}_0(t; J=B) = 1 - \delta_{t,mL/4} ,
\end{equation}
where we used the identity
\begin{equation}
\label{e:Nkpi}
\frac{1}{L} \sum_{k\in \BZ} \sin^2(2k t) = \frac{1 - \delta_{t,mL/4}}{2} ,
\end{equation}
with \(m\) being an integer (\(t=0,1,\ldots\), and \(L\) is even). We find that the entanglement present revivals with a period proportional to the system's size, in which it is maximal during one step. This result is similar to the one obtained for the transverse Ising model \cite{Lakshminarayan-2005}. In the \(J=0\) case, \(\mathcal{Q}_0\) vanishes, while for \(B=0\) it becomes,
\begin{equation}
\label{e:Q0B0}
\mathcal{Q}_0(t; J=0) = 0, \quad \mathcal{Q}_0(t; B=0) = 1 - \cos^4(Jt)  ,
\end{equation}
and tends to \(1/2\) at large times 
\begin{equation}
\label{e:Q0B0p}
\lim_{t \rightarrow \infty} \mathcal{Q}_0(t; B = 0^+) = 5/8  .
\end{equation}
The global entanglement as a function of \(J\) and \(B\) is represented in Fig.~\ref{f:Q0}. On the critical lines \(J=B\), \(\mathcal{Q}_0\) is maximum, reaching the absolute maximum at \(J=B=\pi/2\). Therefore, \(\mathcal{Q}_0\) is a good indicator of the symmetry-breaking phase transition, although it cannot detect the distinct topological phases.

\subsection{Loschmidt echo}

\begin{figure}
  \centering
  \includegraphics[width=1.0\linewidth]{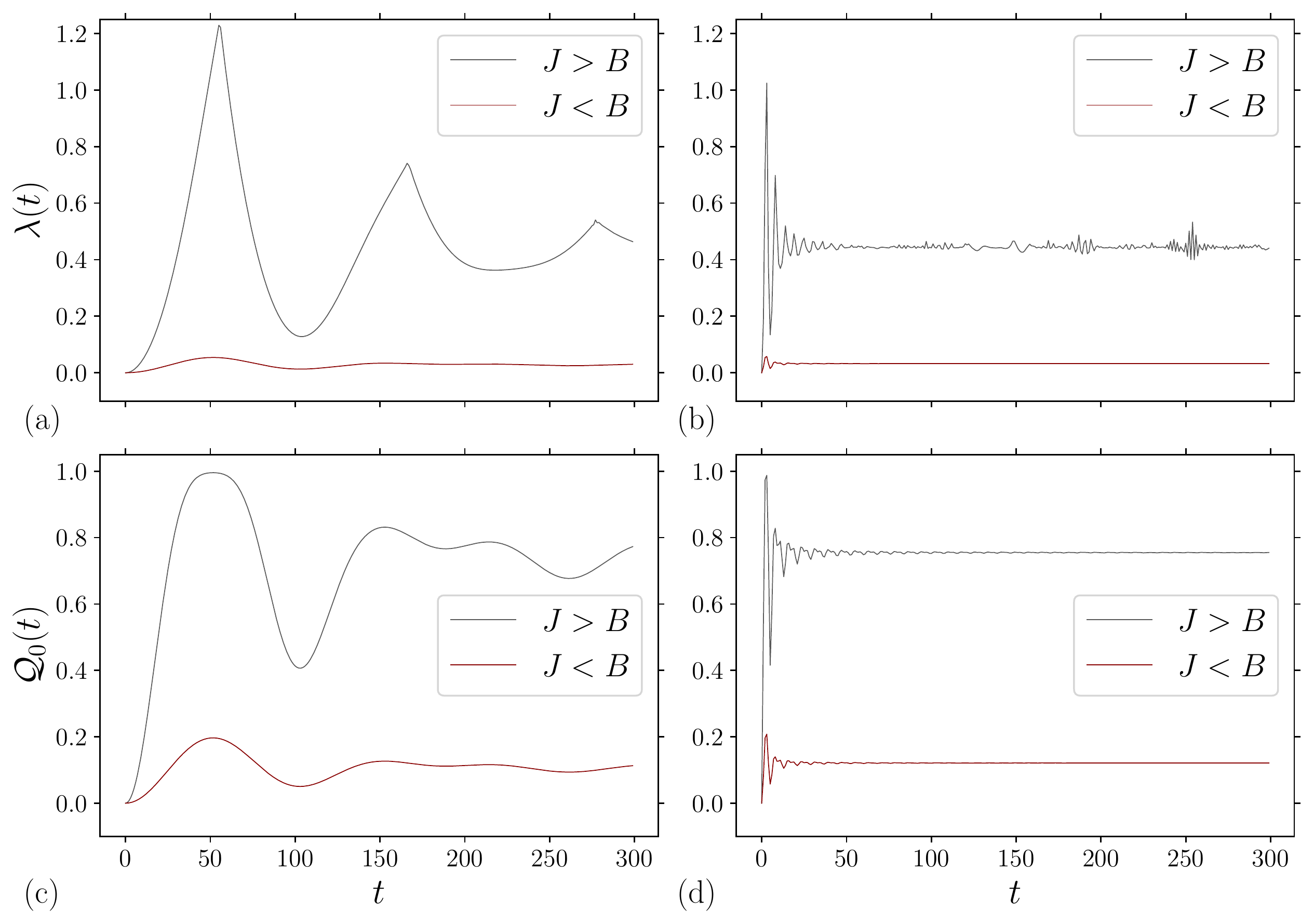}
  \caption{Loschmidt echo rate $\lambda(t)$ [(a) and (b)] and global entanglement $\mathcal{Q}_0(t)$ [(c) and (d)] of the full polarized state. (a) Near $J,B =0$ we find the behavior of the Hamiltonian model; (b) for larger values of $J$ and $B$, $\lambda(t)$ loses regularity. In both (a) and (b), the topological phase ($J>B$, black upper line) is clearly distinguished from the trivial one ($J<B$, red bottom line). The global entanglement [(c) and (d)] follows a similar pattern with higher values for $J>B$.  Parameters: $L=600$, for (a) and (c) ($J=0.03$, $B=0.01$) and ($J=0.01$, $B=0.03$), and for (b) and (d), ($J=0.6$, $B=0.2$) and ($J=0.2$, $B=0.6$).
  \label{f:LE}}
\end{figure}

To characterize the topological phases we investigate the Loschmidt echo \cite{Heyl-2015},
\begin{equation}
\label{e:LE}
\mathcal{L}(t) = |\braket{0 | t}|^2 ,
\end{equation}
here defined for the initial completely polarized pure state \(\ket{0}\), and \(\ket{t}\) is given by \eqref{e:0t}:
\begin{equation}
\label{e:LEt}
\mathcal{L}(t) = \prod_{k \in \BZ} \big\{1 - [1 - n_z(k)^2] \sin^2(\epsilon_k t) \big\} .
\end{equation}
A more suitable quantity is the intensive variable, well defined in the large system limit \(L \rightarrow \infty\), the Loschmidt ratio
\begin{equation}
\label{e:lambda}
\lambda(t) = - \frac{1}{L} \ln \mathcal{L}(t) ,
\end{equation}
which we represent in Fig.~\ref{f:LE}(a) and \ref{f:LE}(b). We observe that the evolution of \(\lambda(t)\) well discriminates the trivial \(J<B\) and nontrivial \(J>B\) phases; however, only near the Hamiltonian limit (in the sense of the Trotter approximation of the evolution operator) is \(\lambda(t)\) smooth, with isolated singularities (cusps) signaling the presence of a dynamical phase transition [Fig.~\ref{f:LE}(a)], similar to the Ising transverse field model \cite{Heyl-2015,Heyl-2018}. For larger values of \((J,B)\), it becomes irregular, with an increase in the frequency of the singularities, and displaying intermittent large fluctuations at long times (revivals) \cite{Happola-2012,Montes-2012,Lakshminarayan-2005}. 

We find that, in Fig.~\ref{f:LE}c, the first peaks of \(\mathcal{Q}_0\) are well correlated with the period \(t=\pi/2J,3\pi/2J,\ldots\approx 52,157,\ldots\), predicted by formula \eqref{e:Q0B0}, valid for vanishing field. Moreover, the long time behavior of both \(\lambda(t)\) and \(\mathcal{Q}_0(t)\) essentially depends on the topological phase rather than the actual values of \(J\) and \(B\).

More precisely, in the small coupling limit \(J,B\rightarrow 0\) with \(J/B \sim O(1)\) one can introduce a small parameter \(\Delta t = t/n\) with \(n\rightarrow\infty\) such that \(J = \bar{J}\Delta t\) and \(B = \bar{B} \Delta t\), where the barred constants are order 1 \(\bar{J}, \bar{B} \sim O(1)\). As a consequence, the Floquet evolution \eqref{e:Flo}, \(F(t) = F(J,B)^t\), can be approximated by \cite{Vanicat-2018}
\begin{equation}
\label{e:trotter}
F(t) \approx \lim_{\Delta t\rightarrow 0} \left( \E^{-\I \bar{H}_C \Delta t} \E^{-\I \bar{H}_B \Delta t} \right)^{ \frac{t}{\Delta t} } = \E^{-\I (\bar{H}_C + \bar{H}_B) t},
\end{equation}
the evolution operator of the transverse cluster model, where, in the barred Hamiltonians, we substituted the barred couplings, \((\bar{J}, \bar{B})\). The behavior of \(\lambda(t)\) [Fig.~\ref{f:LE}(a)] is compatible with the known phenomenology of the model \cite{Montes-2012}. In contrast, when both couplings are order 1, the stroboscopic dynamics appears to be irregular [Fig.~\ref{f:LE}(b), even if the underlying system is integrable]. Note however that the asymptotic levels of the Loschmidt rate (as well as the global entanglement) are comparable in the two regimes.

It is worth noting that the global entanglement given by \eqref{e:Q0t} follows qualitatively the same pattern as the Loschmidt ratio \eqref{e:LEt} [Fig.~\ref{f:LE}(c) and \ref{f:LE}(d)]: The \(J>B\) case corresponds to a high global entanglement with a maximum at the first singular point, and the other case (\(J<B\))  corresponds to a featureless, low-entanglement evolution. However, the $\mathcal{Q}$ measure appears to follow a smooth evolution for small \((J,B)\), in contrast to the appearance of a cusp in \(\lambda\) [Fig.~\ref{f:LE}(a) and \ref{f:LE}(c)], showing that the global entanglement is a poor indicator of the dynamical phase transition. 

In conclusion the integrable Floquet cluster model display a dynamical phase transition between low- and high-entanglement states, extending the phases of the static Hamiltonian's ground state to the driven nonequilibrium regime. The existence of symmetry-protected topological phases in Floquet models is well documented in integrable or near integrable models, for example the ones related to quantum cellular automata \cite{Gopalakrishnan-2018,Friedman-2019} or nonthermal states in constrained systems \cite{Iadecola-2020,Pai-2019,Sellapillay-2022b}; yet it is of interest to investigate Floquet nonergodic states in noiseless interacting systems.

\section{Cluster quantum walk}
\label{S:QW}

To investigate nonergodic behavior beyond the integrable case, we extend the Floquet cluster model to consider the interaction of the chain spins with a moving particle. We introduce then a quantum walk, which in the continuous limit represents a Dirac particle, coupled with the lattice spins by an exchange interaction characterized by the parameter \(J_w\). Related models of interacting quantum walks were used in the study of thermal relaxation \cite{Verga-2019,Verga-2019b} and spin dynamics \cite{Sellapillay-2021}.

The system's total Hilbert space \(\mathcal{H}\) is the tensor product of the walker and spin chain Hilbert spaces; it is then spanned by the basis vectors
\begin{multline}
\label{e:xcs}
\ket{xcs} = \ket{x} \otimes \ket{c} \otimes \ket{s} \in \mathcal{H}, \quad x \in \{0,\ldots,L-1\}, \\ c \in \{0,1\}, \; s \in \{0,\ldots,2^L-1\} ,
\end{multline}
where \(x\) is the particle position, $c$ is the particle spin, which we call the coin state (heads or tails) as usual for quantum walks, and \(s\) is the spin configuration (\(\ket{s} = \ket{s_0 \cdots s_{L-1}}\); cf.\ Sec.~\ref{S:model}). An arbitrary state of the interacting quantum walk can be written in the canonical basis \eqref{e:xcs},
\begin{equation}
\label{e:psixcs}
\ket{\psi} = \sum_{xcs} \psi_{xcs} \ket{xcs}, \quad \sum_{xcs} |\psi_{xcs}|^2 = 1 .
\end{equation}

We choose the particle-spin exchange interaction in the form
\begin{equation}
\label{e:Jex}
W_x(J_w) = \exp\big( \I J_w \tau^{(x)} X_x\big)  ,
\end{equation}
where \(\tau^{(x)}\) is the particle coin operator, the Pauli matrix in the $x$-direction (it flips the coin and acts as the identity on the position and spin spaces). This operator acts at each site \(x\), on the local coin-spin four dimensional Hilbert space, spanned by vectors of the form
\begin{equation}
\label{e:wvec}
\begin{pmatrix} x 0 0_x \\ x 0 1_x \\ x 1 0_x \\ x 1 1_x \end{pmatrix} ,
\end{equation}
where \(0_x,1_x\) denote a spin configuration with a spin up or down at site \(x\), respectively. Note that \(W\) acts as the identity on the position Hilbert subspace. Therefore, \(W_x(J_w)\) couples the coin and the local spin degrees of freedom, entangling the walker with the spin chain.

The motion operator \(M\) of the walker is controlled by its coin degree of freedom \(c\). At each time step we modify the coin state applying a rotation of angle \(\theta\):
\begin{equation}
\label{e:label}
R(\theta) = \exp\big( -\I \theta \tau^{(y)} \big) ,
\end{equation}
whatever the particle position (\(\tau^{(y)}\) is the \(y\) Pauli matrix), followed by a switch of the particle's position between neighbors \(x+1 \rightarrow x\) if the coin state is heads and \(x \rightarrow x+1\) if it is tails: 
\begin{equation}
\label{e:Mop}
M \ket{x 1 s} = \ket{x+1 0 s} , \quad
M \ket{x+1 1 s} = \ket{x 0 s} .
\end{equation}

The sequence \(MC\), where
\begin{equation}
\label{e:Cop}
C(\theta) =  1_L \otimes R(\theta) \otimes 1_{2^L} , 
\end{equation}
is the usual definition of a quantum walk \cite{Kitagawa-2012fk}; with our choice of coin it belongs to the class of Dirac walks, which tend in the continuous limit to the Dirac evolution operator \cite{Meyer-1996sf,Strauch-2006,Di-Molfetta-2012fv,Sellapillay-2021}. Schematically, the walker's behavior in the absence of coupling with the spins (\(J_w=0\)), depends on the rotation angle; when \(\theta\approx\pi/2\) the particle propagates as a chiral excitation walking to the left (if \(c=1\)) or to the right (if \(c=0\)); when \(\theta \approx \pi/4\) the particle propagates ballistically in the two directions.

Finally, we apply the interaction operators, particle-spin \(W\) and spin-spin \(F\). In summary, the one time step operator is
\begin{equation}
\label{e:Fqw}
F_\text{QW} = F(J,B)W(J_w)MC(\theta) , 
\end{equation}
The coupling of the particle with the spins (\(J_w \ne 0\)), through \(W\) and \(F\), introduces an entanglement mechanism between the two parties. It also modifies the system's dispersion properties leading to strong effects on the walker's motion. In Appendix~\ref{S:Aqw} we show the explicit formulas of the different operators in \eqref{e:Fqw}.

It is important to note that the introduction of the coupling \eqref{e:Jex} preserves the \(\mathbb{Z}_2 \times \mathbb{Z}_2\) symmetry of the cluster model, Eq.~\eqref{e:PePo}. However, it deeply modifies the properties of the system, which cannot be consider any more as ``integrable'' \cite{Caux-2011}. One may identify an integrable or quasi-integrable quantum system by the presence of weakly interacting quasiparticles, related to the existence of a set of local conservation laws. The introduction of the walker changes the nature of the degrees of freedom: local and attached to the lattice in the case of the spins, and spread over the whole lattice in the case of the particle. (Appendix~\ref{S:Aqw} shows the complex structure of the Floquet operator, which cannot be associated with a local effective Hamiltonian). 

The nonlocality of the particle wave function allows the spins to indirectly interact at large distance \cite{Klinovaja-2013,Sellapillay-2021}. This fact do not forbids the emergence of dynamical approximated conserved quantities \cite{Haldar-2018,Haldar-2021}, and might facilitate the creation of long range entanglement through the transfer of information carried by the walker to separated locations. Qualitatively, using a naive mean-field reasoning, the spin-particle interaction adds to the applied field \(B\), it might thus suppress its action and restore for example a high entangled phase even if \(J<B\). This is the point we want to study.

In contrast to the effect of an external field, the presence of a self-consistent spin-particle interaction enlarge the cluster model Hilbert space; this means that the state of the spin subsystem is generally mixed. The spin subsystem is then described by the reduced density matrix
\begin{equation}
\label{e:rhos}
\rho_s(t) = \Tr_{xc} \rho(t), \; \rho(t) = \ket{\psi(t)} \bra{\psi(t)}   ,    
\end{equation}
where \(\ket{\psi(t)}\) is the state that has evolved from the initial \(\ket{\psi(0)}\) state, usually a simple product state,
\begin{equation}
\label{e:Fpsi0}
\ket{\psi(t)} = F_\text{QW}^t \ket{\psi(0)} .
\end{equation}
In \eqref{e:rhos} we took the partial trace over the particle and coin degrees of freedom \(\{x,c\}\) of the total system density matrix \(\rho(t)\).

\begin{table}
\caption{\label{t:param} Numerical parameters used in cases (a)-(d) of Fig.~\protect\ref{f:QW}: size $L=14$, particle-spin coupling $J_w=1.6$; $J>B$ [cases (a) and (b)], $J<B$ [cases (c) and (d)]; $\theta \approx \pi/2$ [cases (a) and (c)], $\theta=\pi/4$ [cases (b) and (d)].}
\begin{ruledtabular}
\begin{tabular}{lrrrrr}
Case & $J$ & $B$ & $\theta$ & $\mathcal{Q}_s$ & thermal\\
(a)  & 0.03 & 0.01 & 1.6 & high & no \\
(b)  & 0.03 & 0.01 & $\pi/4$ & low & no \\
(c)  & 0.2 & 0.6 & 1.6 & low & no \\
(d)  & 0.2 & 0.6 & $\pi/4$ & high & yes\\
\end{tabular}
\end{ruledtabular}
\end{table}

Motivated by the formula proposed by Peres \cite{Peres-1984ai}, we introduce a generalization of the Loschmidt overlap to mixed systems in terms of reduced density matrices, here the one of the spin subsystem,
\begin{equation}
\label{e:peres}
\mathcal{L}_s(t) = \frac{\Tr \rho_s(0) \rho_s(t)}{\Tr \rho_s(0)^2} ,\quad
\lambda_s(t) = - \frac{1}{L} \ln \mathcal{L}_s(t) ,
\end{equation}
which, for pure systems and initially mixed systems, reduces to the usual definitions \cite{Peres-1984ai}. Note that this definition does not use the subsystem evolution operator, which in our case is not necessarily a local Floquet Hamiltonian: Formula \eqref{e:peres} generalizes the overlap of two pure states with the one built from the forward and backward time evolution of the partial density matrices of a larger pure system. Alternative definitions do use the subsystem Hamiltonian to evolve an initial mixed state \cite{Heyl-2017,Bhattacharya-2017}.

\begin{figure*}
  \centering%
  \includegraphics[width=0.8\textwidth]{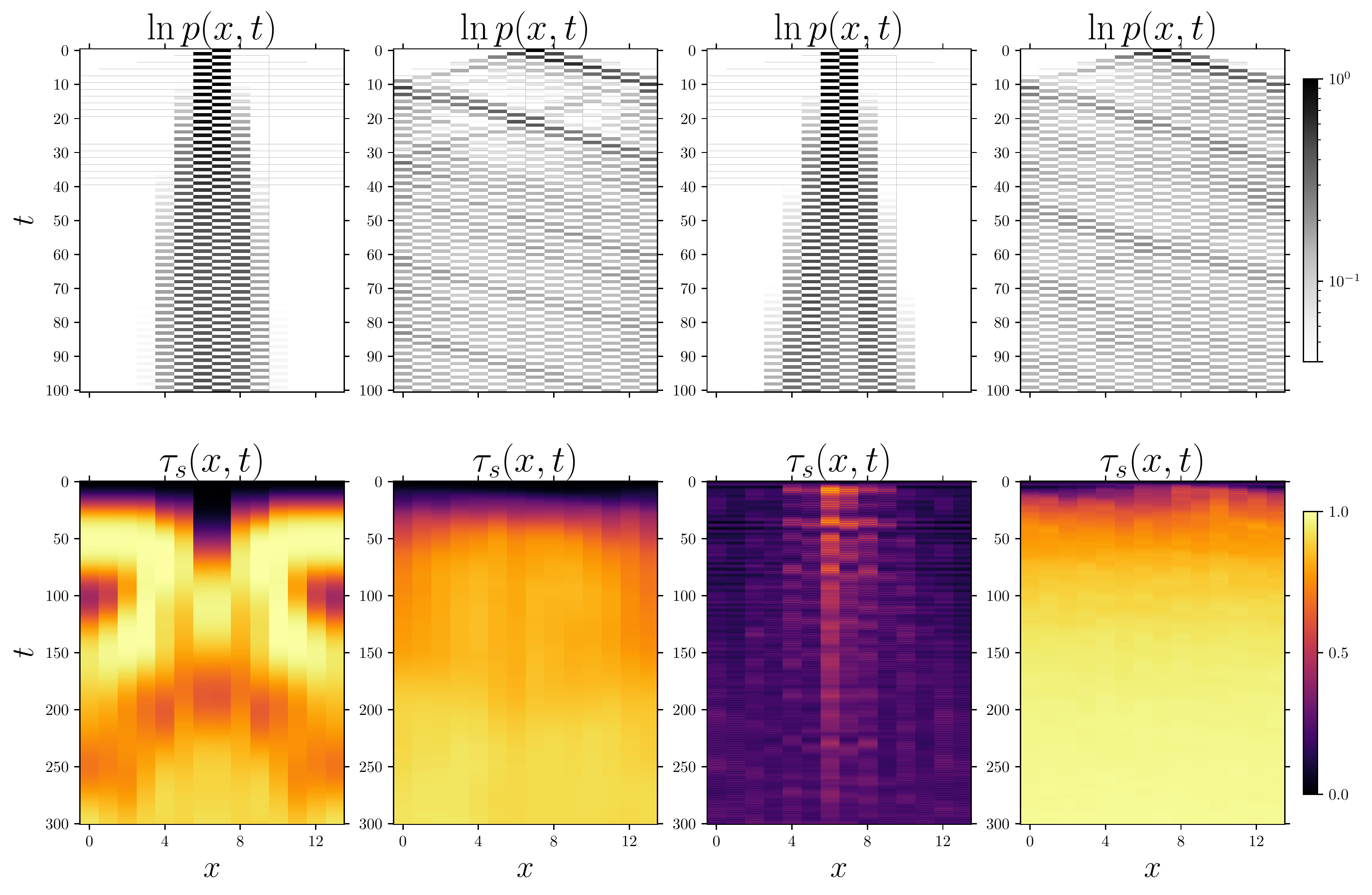}\\
  \includegraphics[width=0.8\textwidth]{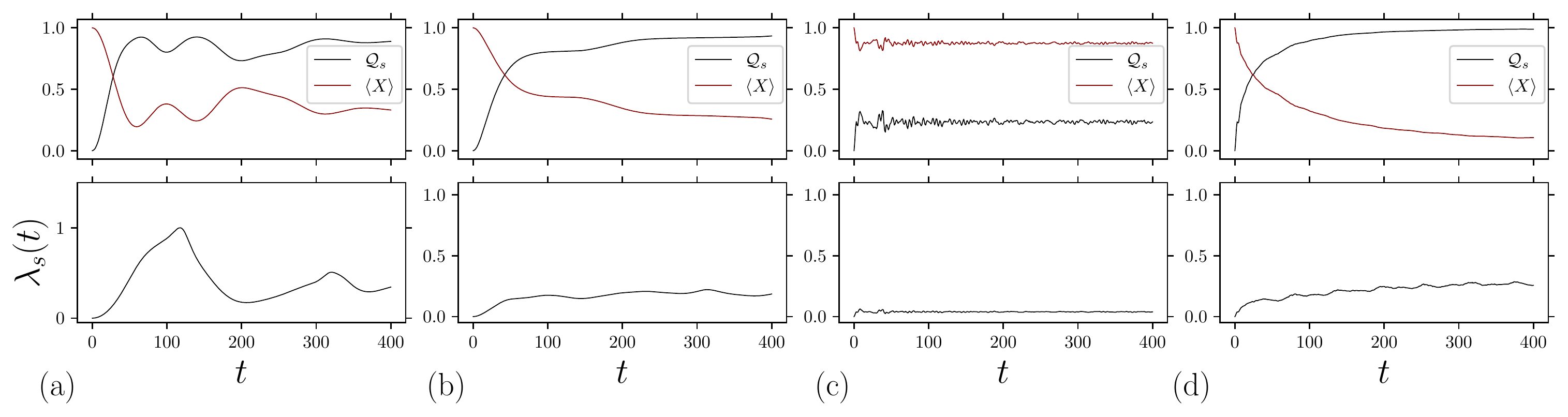}
  \caption{Influence of the spin-particle interaction on the dynamical topological transition. Rows: particle distribution $p$, tangle $\tau$, global spin entanglement $\mathcal{Q}_s(t)$ and magnetization per site $\braket{X}(t)$ , and spin Loschmidt ratio $\lambda_s(t)$; the initial state is $\ket{L/2,0,+}$. For parameters see Table \protect\ref{t:param}. Columns (a), (b) and (d) show evolution towards a highly entangled state, while column (c) show a low-entanglement one. (c) and (d) display a transition between low- and high-entanglement phases induced by the particle (change in the coin angle), even if $J<B$.
  \label{f:QW}}
\end{figure*}

Entanglement of the spin subsystem can be measured with
\begin{equation}
\label{e:Qmix}
\mathcal{Q}_s(t) = \frac{1}{L} \sum_{x=1}^L \tau_s(x,t), \quad
  \tau_s(x,t) = 4 \det \rho_s(x,t),
\end{equation}
where \(\tau_s\) is the so called (one) tangle \cite{Coffman-2000}, and
\[\rho_s(x,t) = \Tr_{\bar{s}_x}\rho_s(t), \quad 
  \bar{s}_x = s_1 \ldots s_{x-1} s_{x+1} \ldots s_L \]
is the density matrix of the spin at site \(x\), which is also a straightforward generalization of the pure state case. For a single spin reduced density matrix, the tangle is useful to visualize the entanglement distribution along the chain. In addition, the walker space density is characterized by the wave function amplitudes \(\psi_{xcs}(t) = \braket{xcs|\psi(t)}\):
\begin{equation}
\label{e:psimix}
p(x,t) = \Tr_{cs} \rho(t) = \sum_{cs} |\psi_{xcs}(t)|^2  .
\end{equation}
It simply gives the probability of finding the walker at position \(x\) and time \(t\), whatever its coin state and surrounding spin configuration. The magnetization distribution per site is defined by
\begin{equation}
\label{e:magn}
\braket{\bm \sigma}(x,t) = \Tr \rho_s(x,t) \bm \sigma ,
\end{equation}
its mean value over the sites, in contrast to the pure state case, does not entirely determines the global entanglement, although in our model where the number of degrees of freedom in the spin subsystem overwhelms the number in the particle subsystem, its behavior is well correlated with the entanglement measures. In particular, in a mixed state, the norm of \(\braket{\bm \sigma}\) is smaller than 1.

We compute the time evolution of an initial state in which all spins are polarized in the \(+\) direction, and the particle is located at \(x = L/2\) with a heads coin (\(c = 0\))
\begin{equation}
\label{e:psi0mix}
\ket{\psi(0)} = \ket{L/2,0}\ket{+}^L .
\end{equation}
In the uncoupled case \(J_w = 0\) this state would evolve into a quantum Dirac walk for the particle subspace \cite{Strauch-2006,Asboth-2012qy} and into the Floquet cluster for the spin subspace (cf.\ Sec.~\ref{S:model}). Instead, when \(J_w \ne 0\), the different degrees of freedom, position, coin and spins, get entangled. We compare the results of four numerical computations labeled as cases (a)-(d) in Fig.~\ref{f:QW}, using the parameters of Table~\ref{t:param}. The four cases use a strong particle-spin coupling \(J_w=1.6\approx \pi/2\), corresponding to the exchange between the coin and local spin states. In addition, we distinguish the \(J>B\) case (a,b), and the \(J<B\) case (c,d). Finally, for the same values of \((J,B)\), we compare the weak dispersive case (a,c), \(\theta \approx \pi/2\), with the strong dispersive one (b,d), \(\theta = \pi/4\).

Case (a) keeps similarities with the integrable case shown in Fig.~\ref{f:LE}(a) and \ref{f:LE}(c), in which a dynamical phase transition appeared. The coupling with the walker results in an increase in the entanglement after the transition, as can be inferred form the behavior of \(\mathcal{Q}_s(t)\) and \(\lambda_s(t)\) [Fig.~\ref{f:QW}(a), rows 3 and 4]. One remarkable effect of the particle-spin interaction is the localization of the walker state around its initial position [Fig.~\ref{f:QW}(a)], row 1). This effect contrasts with the fast dispersion of the \(\theta=\pi/4\) case (b). The asymptotic state reached in both case (a) and case (b), which only differ in the particle dispersion, possesses essentially identical entanglement and magnetization properties. However, the transition dynamics between the initial low- entanglement state and the final highly entangled state is singular in the case where the walker is localized and smooth when the particle-spin interaction is important. It is worth noting that in case (a), the initial growth of the spin entanglement is faster in the region where the particle density is low, and is depleted in the central region as measured by \(\tau_s(x,t)\) [Fig.~\ref{f:QW}(a), row 2].

The effect of the particle on the propagation of the spin's entanglement is also present in cases (c) and (d). In case (c) the walker is localized and the spin entanglement remains inhomogeneous for long times, while its mean value \(\mathcal{Q}_s\) rapidly reaches its saturation value at a low level. The persistence of inhomogeneities in both spin entanglement and magnetization \(\braket{X}\), is typical of nonergodic chaotic states: The evolution of the system and the long time stationary state depend on the initial configuration. Case (d), in which the walker spreads over the whole space, shows entanglement growth and homogenization, with a very small value of the Loschmidt ratio, indicating that the system evolved into a chaotic (thermal) state, in which the magnetization tends to zero after an exponential relaxation. In this \(J<B\) case the transition between the low-entanglement state (c) and high entanglement state (d) is driven by the particle, and controlled by \(\theta\).

A significant difference exists between the integrable case and the interacting quantum walk case. For \(J_w = 0\) the topological phase is magnetically disordered with a vanishing mean magnetization \(\braket{\bm \sigma_x}\); for \(J_w \ne 0\), the topological phase coexists with a magnetic order, as can be verified from the results of Fig.~\ref{f:QW}, where we plot \(\braket{X_x}(t)\) [cf.\ Eq.~\eqref{e:magn}, where \(\rho_s\) replaces \(\ket{\psi}\)]. In the integrable case \(Q_s = 1\) implies \(\braket{X} = 0\), but in the interacting case the spin subsystem can be maximally entangled even in the presence of a finite value of the magnetization, due to the interaction with the particle [Fig.~\ref{f:QW}(a) and \ref{f:QW}(b)]. We further discuss the magnetic order in Appendix~\ref{S:MO}.

In summary (see Table~\ref{t:param} last two columns), case (a) shows a dynamical entanglement transition reminiscent to the one present in the integrable case for \(J>B\); case (b) shows a smooth evolution of the entanglement towards a nonthermal high entangled state; cases (c) and (d) illustrate the transition between a nonthermal chaotic regime and a thermal phase, respectively, induced by the interaction with the particle, and controlled by the quantum walk parameter \(\theta\).

\section{Conclusion}
\label{S:conc}

We investigated the entanglement properties of the Floquet cluster spin chain coupled with a particle via an exchange interaction. The system is invariant with respect to a global \(\mathbb{Z}_2 \times \mathbb{Z}_2\) symmetry. Already in the integrable case, when the coupling vanishes, the periodically driven spin chain exhibits phase transitions between a low-entanglement phase and a topological highly entangled phase, as demonstrated by explicit analytic computations.

More interestingly, we found that the (interacting) combination of the chiral particle motion with the spin chain leads to nonthermal states and dynamical phase transitions between low and high entanglement regimes. In fact, not only can the particle inhibit the topological phase, but also it can be localized by its interaction with the spins; in this case a dynamical phase transition can arise, allowing the initial product state to evolve into a strongly entangled one.

We identified different regimes, some of them extending the properties of the topological ordered cluster phase of the integrable model; however, there is a difference: the emergence of an approximately conserved magnetization. The transition between the initial product state and the high, albeit nonthermal, entangled state could be reached through a dynamical phase transition reminiscent of the one present in the integrable case, or directly following a path of nonexponential relaxation. Exponential relaxation to a paramagnetic regime was also observed, typical of the infinite-temperature phase.

We note that the magnetization vanishes in this paramagnetic regime, even if the applied field is nonzero, and that in the nonthermal phases the presence of a long-time finite magnetization generally correlates with the enhancement of the entanglement with respect to the integrable case. These effects are related to the complex magnetic interactions mediated by the particle scattering off the fixed spins, which can generate effective magnetic (Ising) spin-spin couplings. These magnetic interactions compete with the external field, modifying the original regimes of the effective noninteracting case, and allowing the emergence of other regimes, notably affecting the entanglement dynamics.

In our model the driven frequency is fixed (the unit of time), however it could be of interest to study the behavior of the system as a function of the driving strength, in particular the robustness of the emerging conserved law and the corresponding dynamical phases \cite{Oka-2019,Bukov-2015,Haldar-2022}. Another interesting generalization would be to consider an arbitrary graph of spins, to probe the topological properties of the corresponding cluster state together with those of the underlying graph topology.

The experimental realization of an interacting quantum walk of the type defined by \eqref{e:Fqw} is certainly challenging. However, the basic ingredients (the cluster coupling and the particle walk), can probably be implemented in lattices of Rydberg excited atoms and combined to obtain an effective system approaching \(F_\text{QW}\). Cluster states were recently created using highly selective interactions of atoms in a Rydberg array \cite{Hollerith-2022}. Moreover, it was proposed that with a similar experimental setup, it is possible to simulate a discrete time quantum walk possessing nontrivial topological properties \cite{Khazali-2022}.

In conclusion, ergodicity can then be broken in a periodically driven system by the interplay of qualitatively different interacting degrees of freedom, blocking the evolution towards an infinite temperature state.

\appendix
\section{Eigenvectors and evolution}
\label{S:calc}

In this appendix we give some details about the calculations of Sec.~\ref{S:model}. In terms of the fermion operators \(f_k\), the Fourier components of the Hamiltonian \eqref{e:HF} can be written as
\begin{multline}
\label{e:AHF}
H_k = \epsilon_k \big[ n_z (f^\dagger_k f_k - f_{-k} f^\dagger_{-k}) \\
+ n_- f^\dagger_k f^\dagger_{-k} + n_+ f_{-k} f_{k} \big]  ,
\end{multline}
where \(n_\pm = n_x \pm \I n_y\) and \(n_z\) are given by \eqref{e:nk}. The eigenvectors \eqref{e:kk} are obtained from the following identities:
\begin{gather}
\label{e:Ahk0a}
H_k \ket{0} = \epsilon_k \big( -n_z \ket{0} + n_- f^\dagger_k f^\dagger_{-k} \big)  , \\
\label{e:Ahk0b}
H_k f^\dagger_{\pm k} \ket{0} = 0  , \\
\label{e:Ahk0c}
H_k f^\dagger_k f^\dagger_{-k} \ket{0} = \epsilon_k \big( n_z f^\dagger_{-k} \ket{0} \ket{0} + n_+ \ket{0} \big)  .
\end{gather}
For example, we have
\begin{widetext}
\begin{equation}
\label{e:Ahkkm}
\begin{split}
H_k & \left( n_+ \ket{0} - (1-n_z)  f^\dagger_k f^\dagger_{-k} \ket{0} \right) \\
& = \epsilon_k \left[ -n_z n_+ + n_- n_+  f^\dagger_k f^\dagger_{-k} - (1-n_z) n_z f^\dagger_k f^\dagger_{-k} - (1-n_z) n_+ \right] \ket{0} \\
& = \epsilon_k \left[ -n_+ + (n_x^2 + n_y^2 - n_z + n_z^2) f^\dagger_k f^\dagger_{-k} \right] \ket{0}  , 
\end{split}
\end{equation}
or
\begin{equation}
\label{e:Akkm}
H_k  \left( n_+ - (1-n_z) f^\dagger_k f^\dagger_{-k} \right) \ket{0} = 
 -\epsilon_k \left( n_+ - (1-n_z)  f^\dagger_k f^\dagger_{-k} \right) \ket{0}  , 
\end{equation}
which leads to the eigenvector \(\ket{-kk}\).

The evolution of the vacuum state is
\begin{equation}
\label{e:A0t}
\ket{t} = F(t) \ket{0} = \prod_k F_k(t) \ket{0} .
\end{equation}
We verify that \(\ket{0}\) can be written as a superposition,
\begin{equation}
\label{e:A0sup}
\sqrt{\frac{1-n_z}{2}} \ket{kk} + \frac{n_-}{\sqrt{2(1-n_z)}} \ket{-kk} =
\left( \frac{1-n_z}{2} + \frac{1+n_z}{2} \right) \ket{0} = \ket{0}
\end{equation}
of the \(\pm \epsilon_k\) eigenvectors. Therefore,
\begin{equation}
\label{e:AFk0}
\begin{split}
F_k(t) \ket{0} & = 
\E^{\I \epsilon_k t} \sqrt{\frac{1-n_z}{2}} \ket{kk} + \E^{-\I \epsilon_k t} \frac{n_-}{\sqrt{2(1-n_z)}} \ket{-kk} \\
& = \E^{\I \epsilon_k t} \left( \frac{1-n_z}{2} + \frac{n_-}{2} f^\dagger_{-k} f^\dagger_k \right) \ket{0} + \E^{-\I \epsilon_k t} \left( \frac{1+n_z}{2} - \frac{n_-}{2} f^\dagger_{-k} f^\dagger_k \right) \ket{0} \\
& = \left[ \cos(\epsilon_k t) - \I n_z \sin(\epsilon_k t) + \I n_- \sin(\epsilon_k t) f^\dagger_{-k} f^\dagger_k \right] \ket{0}
\end{split}
\end{equation}
which leads to \eqref{e:0t}. Knowing the explicit expression of \(\ket{t}\) one deduces straightforwardly \(Q_0(t)\) from its definition \eqref{e:Q0t}, as well as \(\lambda(t)\) from \eqref{e:LE} and \eqref{e:lambda}.

A useful formula, used to compute the global entanglement \eqref{e:Q0t}, is the expected value:
\begin{equation}
\label{e:Atfft}
\begin{split}
\braket{t|f_k^\dagger f_k|t} 
&= \bra{0}(c + \I n_z s - \I n_+ s f_{-k} f_k) f^\dagger_k f_k (c - \I n_z s + \I n_- s f_k^\dagger f_{-k}^\dagger)   \ket{0} \\
&= (1-n_z)^2 \sin^2(\epsilon_k t) .
\end{split}
\end{equation}
\end{widetext}
where \(s=\sin(\epsilon_k t)\) and \(c=\cos(\epsilon_k t)\). In the previous equation we used the fact that
\begin{equation}
\label{e:Aft}
f_k \ket{t} = \I n_- \sin(\epsilon_k t) f^\dagger_{-k} \ket{0}  .
\end{equation}
Moreover, the expected value of the spin \(X_x\) does not depend on \(x\) in the state \(\ket{t}\):
\begin{equation}
\label{e:AtXtx}
\braket{t |X_x| t} = 1-\frac{2}{L} \sum_{k_1,k_2} \E^{\I (k_1 - k_2)x} \braket{t|f^\dagger_{k_2} f_{k_1}|t}  ,
\end{equation}
where we used the Fourier transform \eqref{e:Fourier}, because the expected value of the fermion product vanishes for \(k_1 \ne k_2\), leading to the simple relation
\begin{equation}
\label{e:AtXt}
\braket{t |X_x| t} = 1-\frac{2}{L} \sum_{k} \braket{t|f^\dagger_{k} f_{k}|t}  .
\end{equation}
This formula together with \eqref{e:Atfft} allow us to obtain the expression of \(\mathcal{Q}_0\) \eqref{e:Q0t}.

\section{Interacting quantum walk operator}
\label{S:Aqw}

We show the explicit form of the operators \(MC(\theta)\), \(F(J,B)\), and \(W(j_w)\) which define the interacting quantum walk evolution operator \(F_\text{QW}\), \eqref{e:Fqw}.

The combined motion and coin operators governing the particle quantum walk can be written as
\begin{multline}
\label{e:AMC}
MC(\theta) = \sum_x \left[ \ket{x+1}\bra{x} \otimes \ket{0}\bra{1} R(\theta) \right.\\
+ \left. \ket{x}\bra{x+1} \otimes \ket{1}\bra{0} R(\theta) \right] \otimes 1_{2^L}  .
\end{multline}
It acts trivially on the spin Hilbert subspace, and exchange the amplitudes of neighboring sites according to their coin state; the coin state is rotated by an angle \(\theta\).

The cluster Floquet operator \eqref{e:Flo} is readily extended to the whole Hilbert space:
\begin{equation}
\label{e:AFlo}
F(J,B) \rightarrow 1_L \otimes 1_2 \otimes F(J,B) .
\end{equation}

Finally, the coin-spin interaction is given by
\begin{equation}
\label{e:AWJex}
W(J_w) = 1_L \otimes \prod_x \left[ \cos J_w 1_2 \otimes 1_{2^L} + \I \sin J_w \tau^{(x)} \otimes X_x \right]  ,
\end{equation}
where the operator
\begin{equation}
\label{e:AXx}
X_x = 1_{2^{x-1}} \otimes X \otimes 1_{2^{L-x}}  ,
\end{equation}
belongs to the spin Hilbert subspace and \(\tau^{(x)}\) belongs to the coin subspace. It is worth noting that the global structure of \(F_\text{QW}\), in particular due to the sum over \(x\) in \eqref{e:AMC}, cannot be reduced to a tensor product of local unitary operators, as is usual for near-neighbor interacting Hamiltonian systems (as is indeed the case for the driven cluster model). Indeed, if in principle it is possible to embed the quantum walk into a local quantum automaton, it can be done, using a Fourier space representation of the motion operator, at the cost of an exponentially large internal particle space \cite{Brun-2020}; however, in our case, the interaction of the walker, this construction would give a highly nonlocal spin Floquet operator. 

\begin{figure}
  \centering
  \includegraphics[width=1\linewidth]{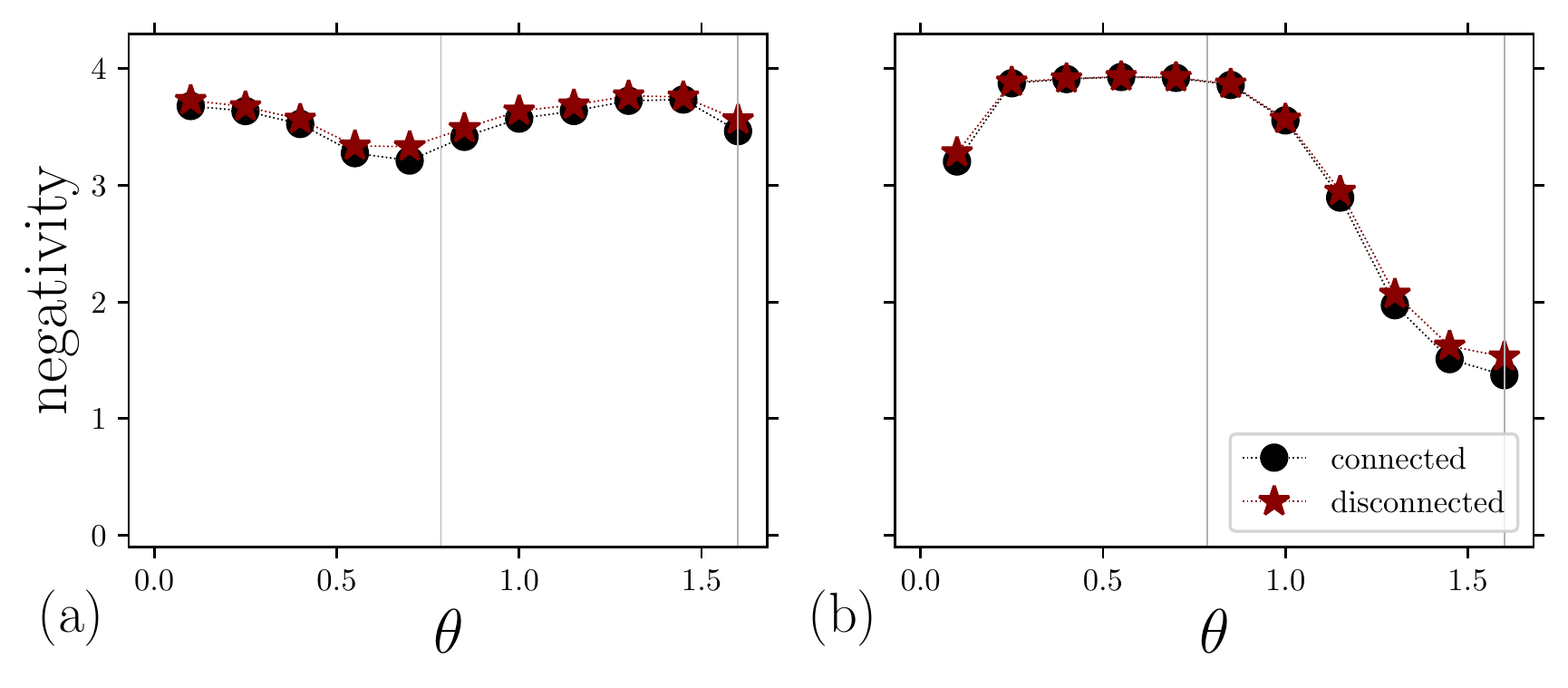}
  \caption{Negativity between two 3-site connected and disconnected sets of spins. Parameters are as in Fig.~\protect\ref{f:QW} (vertical lines at $\theta=\pi/4,1.6$): (a) $J=0.03$, $B=0.01$; (b) $J=0.2$, $B=0.6$; (a) and (b) $J_w=1.6$, $L=12$.
  \label{f:neg}}
\end{figure}

\section{Magnetic order}
\label{S:MO}

We showed in the main text (Sec.~\ref{S:QW}) that a transition was possible even in the \(J<B\) case, between different entanglement regimes, characterized by low-entanglement non ergodic and highly entangled thermal states, whereas in the three-spin interaction dominant case \(J>B\) we observed two types of relaxation toward a highly entangled state, with and without dynamical phase transition. We complement the characterization of these regimes with the computation of the entanglement negativity:
\begin{equation}
\label{e:Nega}
\mathcal{N}_s(A) = \ln |\rho_s^{T_B}|, \quad
  |\rho_s^{T_B}| = 1 + 2 \sum_n |\lambda_n| 
\end{equation}
where AB is a bipartition of the spin subsystem and \(T_B\) denotes the partial transpose over B; the norm of the density matrix is computed from its negative eigenvalues \(\lambda_n < 0\) \cite{Lee-2013}. The negativity is a measure of the entanglement of mixed states, here the spin subsystem, sensitive to the range of entanglement.

\begin{figure}[tb]
  \centering
  \includegraphics[width=1\linewidth]{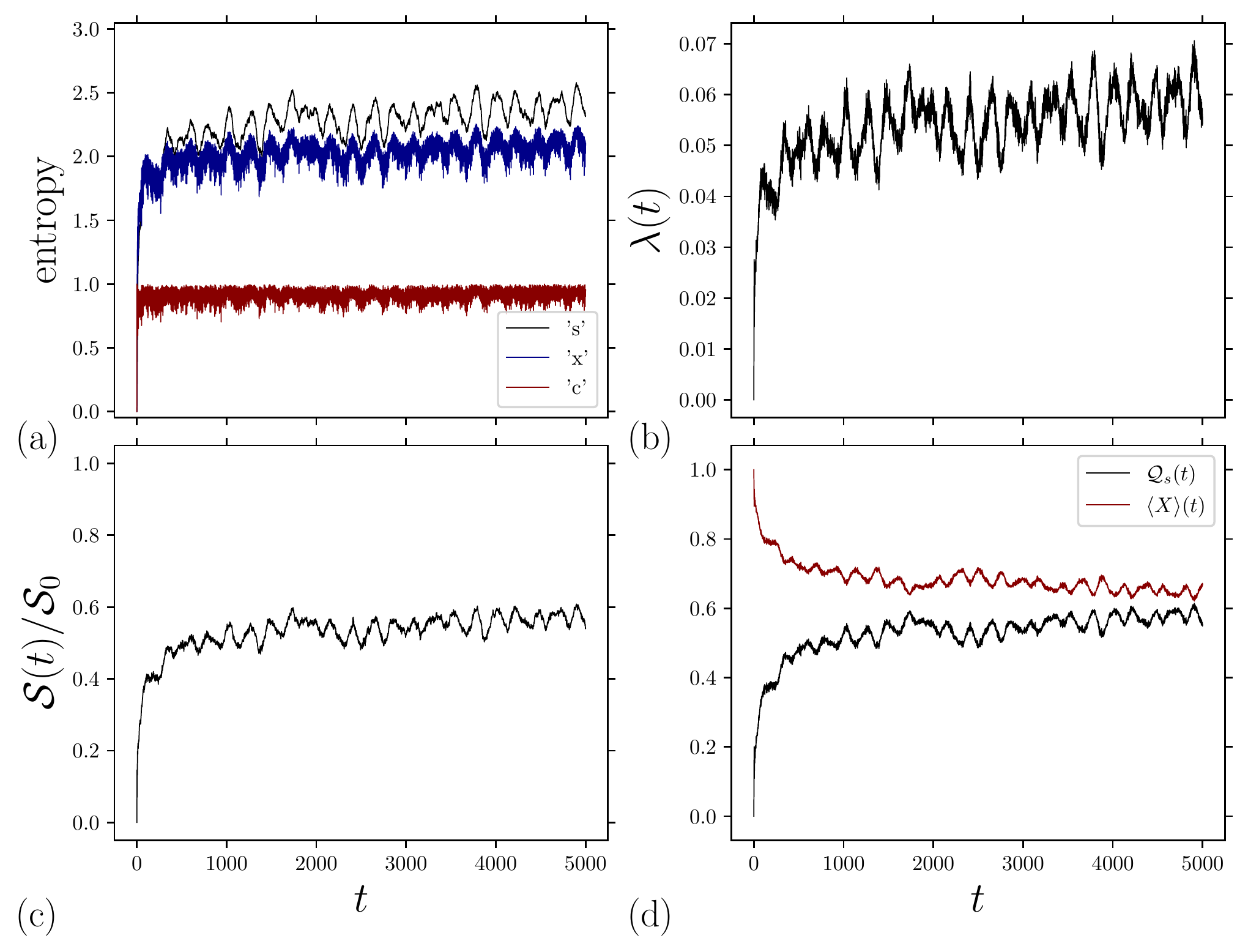}
  \caption{Entanglement in the magnetic ordered phase. (a) Subsystem spin $s$, position $x$, and coin $c$ von Neumann entropies; (b) Loschmidt rate; (c) half-chain entanglement entropy normalized to its maximum value $\mathcal{S}_0 = L/2$; and (d) global entanglement. Parameters: $L=12$, $J=0.2$, $B=0.6$, $J_w=0.5$, and $\theta = \pi/4$ [compare with the parameters of Fig.~\protect\ref{f:QW}(d)].
  \label{f:n_ent}}
\end{figure}

We computed \(\mathcal{N}_s\) as a function the rotation angle \(\theta\), using a connected A set of six spins and a disconnected $3+3$ set (\(L=12\)), for the two regimes of Fig.~\ref{f:QW}, \(J>B\) and \(J<B\). The result is shown in Fig,~\ref{f:neg}. For \(J>B\) the system remains in its high entanglement phase, while for \(J<B\) it displays a transition for values of \(\theta \sim 1.3\) (only the qualitative behavior can be inferred from such small system sizes, here \(L=12\)). A slight difference between the connected and disconnected sets is observed for low entanglement, the disconnected set entanglement being slightly larger than the connected set one.

The parameter \(J_w\) controls the strength of the particle-spin coupling, by tuning its value a near-adiabatic regime in which the fixed spins follow the particle dynamics can be set in. It differs from the topological and paramagnetic phases by its magnetic order. Even in the case where \(J=0\), a two-spin interaction can be mediated by the successive particle scattering off the fixed spins, in much the same way as the Ruderman–Kittel–Kasuya–Yosida (RKKY) interaction of magnetic impurities in a metal(\cite{Kittel-2018}. In this case the orientation of the spins is essentially determined by their indirect interaction through the particle's coin degree of freedom.

Indeed, a simple perturbation expansion argument, in analogy with the above mentioned RKKY interaction \cite{Ruderman-1954}, show that the effective spin Hamiltonian should contain a long range \(XX\) coupling. This coupling is at the origin of the magnetic phase we illustrate in Fig.~\ref{f:n_ent}, which is qualitatively similar to the one in Fig.~\ref{f:QW}(c). To characterize this regime we measured the von Neumann entropy
\begin{equation}
\label{e:vnE}
\mathcal{S}_l(t) = \Tr_{\bar{l}} \ket{\psi(t)}\bra{\psi(t)}, \quad l = \{x,c,s\} ,
\end{equation}
of the \(x,c,s\) subsystems, as well as the half-chain entanglement \(\mathcal{S}(t) \), the global spin entanglement \(\mathcal{Q}_s(t)\), and the expected value of the magnetization vector \(\braket{\bm \sigma_x}\), averaged over \(x\).

\begin{figure}[tb]
  \centering
  \includegraphics[width=0.7\linewidth]{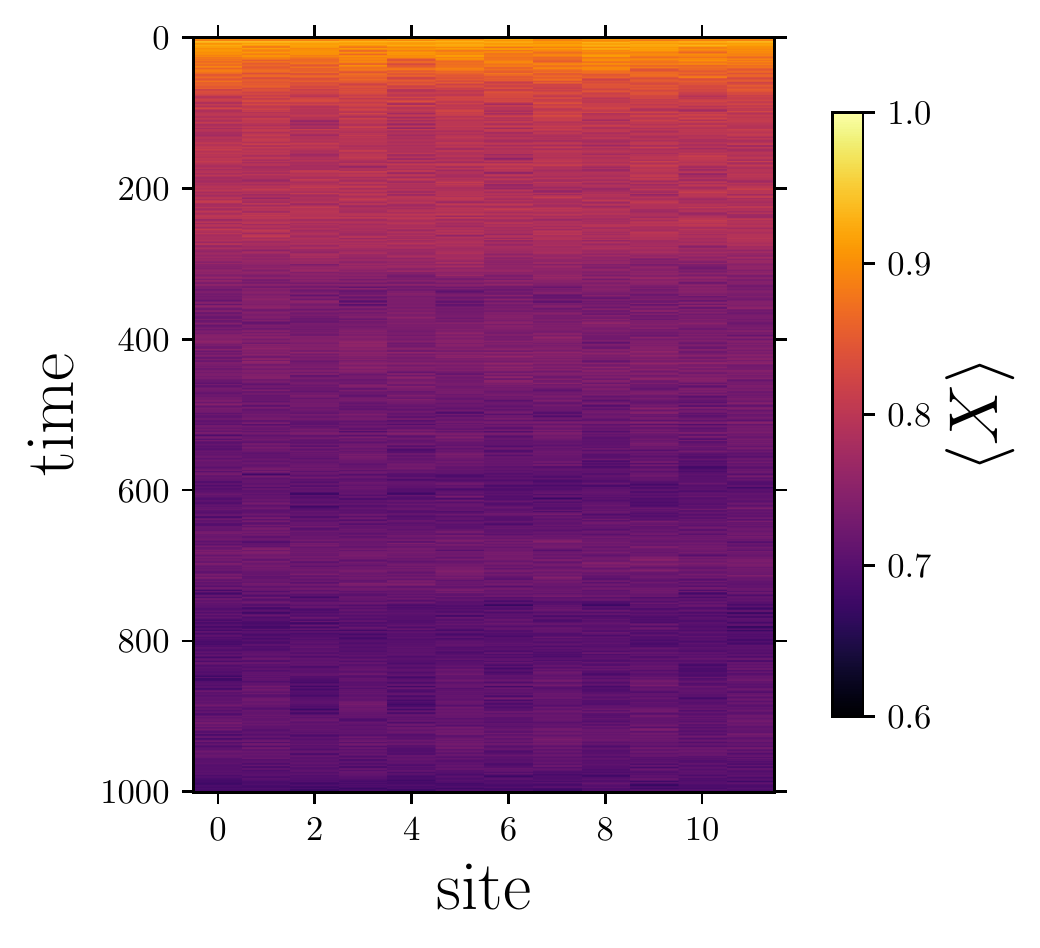}
  \caption{Distribution of the one site magnetization, showing the persistent spatio-temporal fluctuations. Parameters are the sames as in Fig.~\protect\ref{f:n_ent}.
  \label{f:n_sx}}
\end{figure}

We observe that the spin entanglement follows the particle, as measured by the von Neumann entropy [Fig.~\ref{f:n_ent}(a)]; the Loschmidt ratio is small, characteristic of a chaotic state [Fig.~\ref{f:n_ent}(b)]; and the half-chain entropy and the global entanglement [Fig.~\ref{f:n_ent}(c) and \ref{f:n_ent}(d)] saturate at levels well below their maximum values, following the same pattern as the von Neumann entropy, and they also are well correlated with the stochastic variations of the Loschmidt ratio. One important point here is that the stationary state possesses a finite magnetization, signaling a magnetic order (Fig.~\ref{f:n_sx}). This asymptotic magnetization appears as an emergent conserved quantity \cite{Haldar-2018}. We deduce that the magnetic interaction between the chain spins mediated by the walker, in a near adiabatic regime, can establish a high entanglement regime with magnetic order, counterbalancing the paramagnetic effect of the external field (\(B>J\)).

In conclusion, the magnetic order can be attributed to the spin-spin interaction mediated by the walker. It emerges for a range of parameters characterizing the walker motion and coupling with the spins, and its entanglement behavior essentially differs with the paramagnetic (noninteracting) phase.


\begin{thebibliography}{93}%
\makeatletter
\providecommand \@ifxundefined [1]{%
 \@ifx{#1\undefined}
}%
\providecommand \@ifnum [1]{%
 \ifnum #1\expandafter \@firstoftwo
 \else \expandafter \@secondoftwo
 \fi
}%
\providecommand \@ifx [1]{%
 \ifx #1\expandafter \@firstoftwo
 \else \expandafter \@secondoftwo
 \fi
}%
\providecommand \natexlab [1]{#1}%
\providecommand \enquote  [1]{``#1''}%
\providecommand \bibnamefont  [1]{#1}%
\providecommand \bibfnamefont [1]{#1}%
\providecommand \citenamefont [1]{#1}%
\providecommand \href@noop [0]{\@secondoftwo}%
\providecommand \href [0]{\begingroup \@sanitize@url \@href}%
\providecommand \@href[1]{\@@startlink{#1}\@@href}%
\providecommand \@@href[1]{\endgroup#1\@@endlink}%
\providecommand \@sanitize@url [0]{\catcode `\\12\catcode `\$12\catcode
  `\&12\catcode `\#12\catcode `\^12\catcode `\_12\catcode `\%12\relax}%
\providecommand \@@startlink[1]{}%
\providecommand \@@endlink[0]{}%
\providecommand \url  [0]{\begingroup\@sanitize@url \@url }%
\providecommand \@url [1]{\endgroup\@href {#1}{\urlprefix }}%
\providecommand \urlprefix  [0]{URL }%
\providecommand \Eprint [0]{\href }%
\providecommand \doibase [0]{https://doi.org/}%
\providecommand \selectlanguage [0]{\@gobble}%
\providecommand \bibinfo  [0]{\@secondoftwo}%
\providecommand \bibfield  [0]{\@secondoftwo}%
\providecommand \translation [1]{[#1]}%
\providecommand \BibitemOpen [0]{}%
\providecommand \bibitemStop [0]{}%
\providecommand \bibitemNoStop [0]{.\EOS\space}%
\providecommand \EOS [0]{\spacefactor3000\relax}%
\providecommand \BibitemShut  [1]{\csname bibitem#1\endcsname}%
\let\auto@bib@innerbib\@empty
%</preamble>
\bibitem [{\citenamefont {Doherty}\ and\ \citenamefont
  {Bartlett}(2009)}]{Doherty-2009}%
  \BibitemOpen
  \bibfield  {author} {\bibinfo {author} {\bibfnamefont {A.~C.}\ \bibnamefont
  {Doherty}}\ and\ \bibinfo {author} {\bibfnamefont {S.~D.}\ \bibnamefont
  {Bartlett}},\ }\bibfield  {title} {\bibinfo {title} {Identifying {{Phases}}
  of {{Quantum Many-Body Systems That Are Universal}} for {{Quantum
  Computation}}},\ }\href {https://doi.org/10.1103/PhysRevLett.103.020506}
  {\bibfield  {journal} {\bibinfo  {journal} {Phys. Rev. Lett.}\ }\textbf
  {\bibinfo {volume} {103}},\ \bibinfo {pages} {020506} (\bibinfo {year}
  {2009})}\BibitemShut {NoStop}%
\bibitem [{\citenamefont {Stephen}\ \emph {et~al.}(2019)\citenamefont
  {Stephen}, \citenamefont {Nautrup}, \citenamefont {{Bermejo-Vega}},
  \citenamefont {Eisert},\ and\ \citenamefont {Raussendorf}}]{Stephen-2019}%
  \BibitemOpen
  \bibfield  {author} {\bibinfo {author} {\bibfnamefont {D.~T.}\ \bibnamefont
  {Stephen}}, \bibinfo {author} {\bibfnamefont {H.~P.}\ \bibnamefont
  {Nautrup}}, \bibinfo {author} {\bibfnamefont {J.}~\bibnamefont
  {{Bermejo-Vega}}}, \bibinfo {author} {\bibfnamefont {J.}~\bibnamefont
  {Eisert}},\ and\ \bibinfo {author} {\bibfnamefont {R.}~\bibnamefont
  {Raussendorf}},\ }\bibfield  {title} {\bibinfo {title} {Subsystem symmetries,
  quantum cellular automata, and computational phases of quantum matter},\
  }\href {https://doi.org/10.22331/q-2019-05-20-142} {\bibfield  {journal}
  {\bibinfo  {journal} {Quantum}\ }\textbf {\bibinfo {volume} {3}},\ \bibinfo
  {pages} {142} (\bibinfo {year} {2019})}\BibitemShut {NoStop}%
\bibitem [{\citenamefont {Shor}(1996)}]{Shor-1996}%
  \BibitemOpen
  \bibfield  {author} {\bibinfo {author} {\bibfnamefont {P.~W.}\ \bibnamefont
  {Shor}},\ }\bibfield  {title} {\bibinfo {title} {Fault-tolerant quantum
  computation},\ }in\ \href {https://doi.org/10.1109/SFCS.1996.548464} {\emph
  {\bibinfo {booktitle} {Proc. 37th {{Conf}}. {{Found}}. {{Comput}}.
  {{Sci}}.}}}\ (\bibinfo  {publisher} {{IEEE}},\ \bibinfo {year} {1996})\ pp.\
  \bibinfo {pages} {56--65}\BibitemShut {NoStop}%
\bibitem [{\citenamefont {Preskill}(1998)}]{Preskill-1998}%
  \BibitemOpen
  \bibfield  {author} {\bibinfo {author} {\bibfnamefont {J.}~\bibnamefont
  {Preskill}},\ }\bibfield  {title} {\bibinfo {title} {Fault-tolerant quantum
  computation},\ }in\ \href {https://doi.org/10.1142/9789812385253_0008} {\emph
  {\bibinfo {booktitle} {Introduction to {{Quantum Computation}} and
  {{Information}}}}}\ (\bibinfo  {publisher} {{World Scientific}},\ \bibinfo
  {address} {{Singapore}},\ \bibinfo {year} {1998})\ pp.\ \bibinfo {pages}
  {213--269}\BibitemShut {NoStop}%
\bibitem [{\citenamefont {Gottesman}(1997)}]{Gottesman-1997}%
  \BibitemOpen
  \bibfield  {author} {\bibinfo {author} {\bibfnamefont {D.}~\bibnamefont
  {Gottesman}},\ }\bibfield  {title} {\bibinfo {title} {Stabilizer {{Codes}}
  and {{Quantum Error Correction}}},\ }\href
  {http://arxiv.org/abs/quant-ph/9705052} {\bibfield  {journal} {\bibinfo
  {journal} {ArXivquant-Ph9705052}\ } (\bibinfo {year} {1997})},\ \Eprint
  {https://arxiv.org/abs/quant-ph/9705052} {arXiv:quant-ph/9705052}
  \BibitemShut {NoStop}%
\bibitem [{\citenamefont {Kitaev}(2003)}]{Kitaev-2003fk}%
  \BibitemOpen
  \bibfield  {author} {\bibinfo {author} {\bibfnamefont {A.~Y.}\ \bibnamefont
  {Kitaev}},\ }\bibfield  {title} {\bibinfo {title} {Fault-tolerant quantum
  computation by anyons},\ }\href
  {https://doi.org/10.1016/S0003-4916(02)00018-0} {\bibfield  {journal}
  {\bibinfo  {journal} {Ann. Phys.}\ }\textbf {\bibinfo {volume} {303}},\
  \bibinfo {pages} {2} (\bibinfo {year} {2003})}\BibitemShut {NoStop}%
\bibitem [{\citenamefont {Briegel}\ and\ \citenamefont
  {Raussendorf}(2001)}]{Briegel-2001fk}%
  \BibitemOpen
  \bibfield  {author} {\bibinfo {author} {\bibfnamefont {H.~J.}\ \bibnamefont
  {Briegel}}\ and\ \bibinfo {author} {\bibfnamefont {R.}~\bibnamefont
  {Raussendorf}},\ }\bibfield  {title} {\bibinfo {title} {Persistent
  {{Entanglement}} in {{Arrays}} of {{Interacting Particles}}},\ }\href
  {https://doi.org/10.1103/PhysRevLett.86.910} {\bibfield  {journal} {\bibinfo
  {journal} {Phys. Rev. Lett.}\ }\textbf {\bibinfo {volume} {86}},\ \bibinfo
  {pages} {910} (\bibinfo {year} {2001})}\BibitemShut {NoStop}%
\bibitem [{\citenamefont {Freedman}\ \emph {et~al.}(2003)\citenamefont
  {Freedman}, \citenamefont {Kitaev}, \citenamefont {Larsen},\ and\
  \citenamefont {Wang}}]{Freedman-2003}%
  \BibitemOpen
  \bibfield  {author} {\bibinfo {author} {\bibfnamefont {M.}~\bibnamefont
  {Freedman}}, \bibinfo {author} {\bibfnamefont {A.}~\bibnamefont {Kitaev}},
  \bibinfo {author} {\bibfnamefont {M.}~\bibnamefont {Larsen}},\ and\ \bibinfo
  {author} {\bibfnamefont {Z.}~\bibnamefont {Wang}},\ }\bibfield  {title}
  {\bibinfo {title} {Topological quantum computation},\ }\href
  {https://doi.org/10.1090/S0273-0979-02-00964-3} {\bibfield  {journal}
  {\bibinfo  {journal} {Bull. Amer. Math. Soc.}\ }\textbf {\bibinfo {volume}
  {40}},\ \bibinfo {pages} {31} (\bibinfo {year} {2003})}\BibitemShut {NoStop}%
\bibitem [{\citenamefont {Raussendorf}\ and\ \citenamefont
  {Briegel}(2001)}]{Raussendorf-2001uq}%
  \BibitemOpen
  \bibfield  {author} {\bibinfo {author} {\bibfnamefont {R.}~\bibnamefont
  {Raussendorf}}\ and\ \bibinfo {author} {\bibfnamefont {H.~J.}\ \bibnamefont
  {Briegel}},\ }\bibfield  {title} {\bibinfo {title} {A {{One-Way Quantum
  Computer}}},\ }\href {https://doi.org/10.1103/PhysRevLett.86.5188} {\bibfield
   {journal} {\bibinfo  {journal} {Phys. Rev. Lett.}\ }\textbf {\bibinfo
  {volume} {86}},\ \bibinfo {pages} {5188} (\bibinfo {year}
  {2001})}\BibitemShut {NoStop}%
\bibitem [{\citenamefont {Dennis}\ \emph {et~al.}(2002)\citenamefont {Dennis},
  \citenamefont {Kitaev}, \citenamefont {Landahl},\ and\ \citenamefont
  {Preskill}}]{Dennis-2002}%
  \BibitemOpen
  \bibfield  {author} {\bibinfo {author} {\bibfnamefont {E.}~\bibnamefont
  {Dennis}}, \bibinfo {author} {\bibfnamefont {A.}~\bibnamefont {Kitaev}},
  \bibinfo {author} {\bibfnamefont {A.}~\bibnamefont {Landahl}},\ and\ \bibinfo
  {author} {\bibfnamefont {J.}~\bibnamefont {Preskill}},\ }\bibfield  {title}
  {\bibinfo {title} {Topological quantum memory},\ }\href
  {https://doi.org/10.1063/1.1499754} {\bibfield  {journal} {\bibinfo
  {journal} {Journal of Mathematical Physics}\ }\textbf {\bibinfo {volume}
  {43}},\ \bibinfo {pages} {4452} (\bibinfo {year} {2002})}\BibitemShut
  {NoStop}%
\bibitem [{\citenamefont {Brown}\ \emph {et~al.}(2016)\citenamefont {Brown},
  \citenamefont {Loss}, \citenamefont {Pachos}, \citenamefont {Self},\ and\
  \citenamefont {Wootton}}]{Brown-2016}%
  \BibitemOpen
  \bibfield  {author} {\bibinfo {author} {\bibfnamefont {B.~J.}\ \bibnamefont
  {Brown}}, \bibinfo {author} {\bibfnamefont {D.}~\bibnamefont {Loss}},
  \bibinfo {author} {\bibfnamefont {J.~K.}\ \bibnamefont {Pachos}}, \bibinfo
  {author} {\bibfnamefont {C.~N.}\ \bibnamefont {Self}},\ and\ \bibinfo
  {author} {\bibfnamefont {J.~R.}\ \bibnamefont {Wootton}},\ }\bibfield
  {title} {\bibinfo {title} {Quantum memories at finite temperature},\ }\href
  {https://doi.org/10.1103/RevModPhys.88.045005} {\bibfield  {journal}
  {\bibinfo  {journal} {Rev. Mod. Phys.}\ }\textbf {\bibinfo {volume} {88}},\
  \bibinfo {pages} {045005} (\bibinfo {year} {2016})}\BibitemShut {NoStop}%
\bibitem [{\citenamefont {Roberts}(2019)}]{Roberts-2019}%
  \BibitemOpen
  \bibfield  {author} {\bibinfo {author} {\bibfnamefont {S.}~\bibnamefont
  {Roberts}},\ }\emph {\bibinfo {title} {Symmetry-{{Protected Topological
  Phases}} for {{Robust Quantum Computation}}}},\ \href
  {https://ses.library.usyd.edu.au/handle/2123/21192} {\bibinfo {type}
  {Thesis}} (\bibinfo {year} {2019})\BibitemShut {NoStop}%
\bibitem [{\citenamefont {Wildeboer}\ \emph {et~al.}(2022)\citenamefont
  {Wildeboer}, \citenamefont {Iadecola},\ and\ \citenamefont
  {Williamson}}]{Wildeboer-2022}%
  \BibitemOpen
  \bibfield  {author} {\bibinfo {author} {\bibfnamefont {J.}~\bibnamefont
  {Wildeboer}}, \bibinfo {author} {\bibfnamefont {T.}~\bibnamefont
  {Iadecola}},\ and\ \bibinfo {author} {\bibfnamefont {D.~J.}\ \bibnamefont
  {Williamson}},\ }\bibfield  {title} {\bibinfo {title} {Symmetry-{{Protected
  Infinite-Temperature Quantum Memory}} from {{Subsystem Codes}}},\ }\href
  {https://doi.org/10.1103/PRXQuantum.3.020330} {\bibfield  {journal} {\bibinfo
   {journal} {PRX Quantum}\ }\textbf {\bibinfo {volume} {3}},\ \bibinfo {pages}
  {020330} (\bibinfo {year} {2022})}\BibitemShut {NoStop}%
\bibitem [{\citenamefont {D'Alessio}\ \emph {et~al.}(2016)\citenamefont
  {D'Alessio}, \citenamefont {Kafri}, \citenamefont {Polkovnikov},\ and\
  \citenamefont {Rigol}}]{Alessio-2016fj}%
  \BibitemOpen
  \bibfield  {author} {\bibinfo {author} {\bibfnamefont {L.}~\bibnamefont
  {D'Alessio}}, \bibinfo {author} {\bibfnamefont {Y.}~\bibnamefont {Kafri}},
  \bibinfo {author} {\bibfnamefont {A.}~\bibnamefont {Polkovnikov}},\ and\
  \bibinfo {author} {\bibfnamefont {M.}~\bibnamefont {Rigol}},\ }\bibfield
  {title} {\bibinfo {title} {From quantum chaos and eigenstate thermalization
  to statistical mechanics and thermodynamics},\ }\href
  {https://doi.org/10.1080/00018732.2016.1198134} {\bibfield  {journal}
  {\bibinfo  {journal} {Adv. Phys.}\ }\textbf {\bibinfo {volume} {65}},\
  \bibinfo {pages} {239} (\bibinfo {year} {2016})}\BibitemShut {NoStop}%
\bibitem [{\citenamefont {Gross}\ \emph {et~al.}(2009)\citenamefont {Gross},
  \citenamefont {Flammia},\ and\ \citenamefont {Eisert}}]{Gross-2009uq}%
  \BibitemOpen
  \bibfield  {author} {\bibinfo {author} {\bibfnamefont {D.}~\bibnamefont
  {Gross}}, \bibinfo {author} {\bibfnamefont {S.~T.}\ \bibnamefont {Flammia}},\
  and\ \bibinfo {author} {\bibfnamefont {J.}~\bibnamefont {Eisert}},\
  }\bibfield  {title} {\bibinfo {title} {Most {{Quantum States Are Too
  Entangled To Be Useful As Computational Resources}}},\ }\href
  {https://doi.org/10.1103/PhysRevLett.102.190501} {\bibfield  {journal}
  {\bibinfo  {journal} {Phys. Rev. Lett.}\ }\textbf {\bibinfo {volume} {102}},\
  \bibinfo {pages} {190501} (\bibinfo {year} {2009})}\BibitemShut {NoStop}%
\bibitem [{\citenamefont {D’Alessio}\ and\ \citenamefont
  {Rigol}(2014)}]{DAlessio-2014}%
  \BibitemOpen
  \bibfield  {author} {\bibinfo {author} {\bibfnamefont {L.}~\bibnamefont
  {D’Alessio}}\ and\ \bibinfo {author} {\bibfnamefont {M.}~\bibnamefont
  {Rigol}},\ }\bibfield  {title} {\bibinfo {title} {Long-time {{Behavior}} of
  {{Isolated Periodically Driven Interacting Lattice Systems}}},\ }\href
  {https://doi.org/10.1103/PhysRevX.4.041048} {\bibfield  {journal} {\bibinfo
  {journal} {Phys. Rev. X}\ }\textbf {\bibinfo {volume} {4}},\ \bibinfo {pages}
  {041048} (\bibinfo {year} {2014})}\BibitemShut {NoStop}%
\bibitem [{\citenamefont {Lazarides}\ \emph {et~al.}(2014)\citenamefont
  {Lazarides}, \citenamefont {Das},\ and\ \citenamefont
  {Moessner}}]{Lazarides-2014}%
  \BibitemOpen
  \bibfield  {author} {\bibinfo {author} {\bibfnamefont {A.}~\bibnamefont
  {Lazarides}}, \bibinfo {author} {\bibfnamefont {A.}~\bibnamefont {Das}},\
  and\ \bibinfo {author} {\bibfnamefont {R.}~\bibnamefont {Moessner}},\
  }\bibfield  {title} {\bibinfo {title} {Equilibrium states of generic quantum
  systems subject to periodic driving},\ }\href
  {https://doi.org/10.1103/PhysRevE.90.012110} {\bibfield  {journal} {\bibinfo
  {journal} {Phys. Rev. E}\ }\textbf {\bibinfo {volume} {90}},\ \bibinfo
  {pages} {012110} (\bibinfo {year} {2014})}\BibitemShut {NoStop}%
\bibitem [{\citenamefont {Chamon}(2005)}]{Chamon-2005}%
  \BibitemOpen
  \bibfield  {author} {\bibinfo {author} {\bibfnamefont {C.}~\bibnamefont
  {Chamon}},\ }\bibfield  {title} {\bibinfo {title} {Quantum {{Glassiness}} in
  {{Strongly Correlated Clean Systems}}: {{An Example}} of {{Topological
  Overprotection}}},\ }\href {https://doi.org/10.1103/PhysRevLett.94.040402}
  {\bibfield  {journal} {\bibinfo  {journal} {Phys. Rev. Lett.}\ }\textbf
  {\bibinfo {volume} {94}},\ \bibinfo {pages} {040402} (\bibinfo {year}
  {2005})}\BibitemShut {NoStop}%
\bibitem [{\citenamefont {Sala}\ \emph {et~al.}(2020)\citenamefont {Sala},
  \citenamefont {Rakovszky}, \citenamefont {Verresen}, \citenamefont {Knap},\
  and\ \citenamefont {Pollmann}}]{Sala-2020}%
  \BibitemOpen
  \bibfield  {author} {\bibinfo {author} {\bibfnamefont {P.}~\bibnamefont
  {Sala}}, \bibinfo {author} {\bibfnamefont {T.}~\bibnamefont {Rakovszky}},
  \bibinfo {author} {\bibfnamefont {R.}~\bibnamefont {Verresen}}, \bibinfo
  {author} {\bibfnamefont {M.}~\bibnamefont {Knap}},\ and\ \bibinfo {author}
  {\bibfnamefont {F.}~\bibnamefont {Pollmann}},\ }\bibfield  {title} {\bibinfo
  {title} {Ergodicity {{Breaking Arising}} from {{Hilbert Space Fragmentation}}
  in {{Dipole-Conserving Hamiltonians}}},\ }\href
  {https://doi.org/10.1103/PhysRevX.10.011047} {\bibfield  {journal} {\bibinfo
  {journal} {Phys. Rev. X}\ }\textbf {\bibinfo {volume} {10}},\ \bibinfo
  {pages} {011047} (\bibinfo {year} {2020})}\BibitemShut {NoStop}%
\bibitem [{\citenamefont {Scherg}\ \emph {et~al.}(2021)\citenamefont {Scherg},
  \citenamefont {Kohlert}, \citenamefont {Sala}, \citenamefont {Pollmann},
  \citenamefont {Hebbe~Madhusudhana}, \citenamefont {Bloch},\ and\
  \citenamefont {Aidelsburger}}]{Scherg-2021a}%
  \BibitemOpen
  \bibfield  {author} {\bibinfo {author} {\bibfnamefont {S.}~\bibnamefont
  {Scherg}}, \bibinfo {author} {\bibfnamefont {T.}~\bibnamefont {Kohlert}},
  \bibinfo {author} {\bibfnamefont {P.}~\bibnamefont {Sala}}, \bibinfo {author}
  {\bibfnamefont {F.}~\bibnamefont {Pollmann}}, \bibinfo {author}
  {\bibfnamefont {B.}~\bibnamefont {Hebbe~Madhusudhana}}, \bibinfo {author}
  {\bibfnamefont {I.}~\bibnamefont {Bloch}},\ and\ \bibinfo {author}
  {\bibfnamefont {M.}~\bibnamefont {Aidelsburger}},\ }\bibfield  {title}
  {\bibinfo {title} {Observing non-ergodicity due to kinetic constraints in
  tilted {{Fermi-Hubbard}} chains},\ }\href
  {https://doi.org/10.1038/s41467-021-24726-0} {\bibfield  {journal} {\bibinfo
  {journal} {Nat Commun}\ }\textbf {\bibinfo {volume} {12}},\ \bibinfo {pages}
  {4490} (\bibinfo {year} {2021})}\BibitemShut {NoStop}%
\bibitem [{\citenamefont {Shiraishi}\ and\ \citenamefont
  {Mori}(2017)}]{Shiraishi-2017}%
  \BibitemOpen
  \bibfield  {author} {\bibinfo {author} {\bibfnamefont {N.}~\bibnamefont
  {Shiraishi}}\ and\ \bibinfo {author} {\bibfnamefont {T.}~\bibnamefont
  {Mori}},\ }\bibfield  {title} {\bibinfo {title} {Systematic {{Construction}}
  of {{Counterexamples}} to the {{Eigenstate Thermalization Hypothesis}}},\
  }\href {https://doi.org/10.1103/PhysRevLett.119.030601} {\bibfield  {journal}
  {\bibinfo  {journal} {Phys. Rev. Lett.}\ }\textbf {\bibinfo {volume} {119}},\
  \bibinfo {pages} {030601} (\bibinfo {year} {2017})}\BibitemShut {NoStop}%
\bibitem [{\citenamefont {Papić}(2022)}]{Papic-2022}%
  \BibitemOpen
  \bibfield  {author} {\bibinfo {author} {\bibfnamefont {Z.}~\bibnamefont
  {Papić}},\ }\bibfield  {title} {\bibinfo {title} {Weak {{Ergodicity Breaking
  Through}} the {{Lens}} of {{Quantum Entanglement}}},\ }in\ \href
  {https://doi.org/10.1007/978-3-031-03998-0_13} {\emph {\bibinfo {booktitle}
  {Entanglement in {{Spin Chains}}: {{From Theory}} to {{Quantum Technology
  Applications}}}}},\ \bibinfo {series and number} {Quantum {{Science}} and
  {{Technology}}},\ \bibinfo {editor} {edited by\ \bibinfo {editor}
  {\bibfnamefont {A.}~\bibnamefont {Bayat}}, \bibinfo {editor} {\bibfnamefont
  {S.}~\bibnamefont {Bose}},\ and\ \bibinfo {editor} {\bibfnamefont
  {H.}~\bibnamefont {Johannesson}}}\ (\bibinfo  {publisher} {{Springer}},\
  \bibinfo {address} {{Cham, Switzerland}},\ \bibinfo {year} {2022})\ pp.\
  \bibinfo {pages} {341--395}\BibitemShut {NoStop}%
\bibitem [{\citenamefont {Gopalakrishnan}\ and\ \citenamefont
  {Zakirov}(2018)}]{Gopalakrishnan-2018}%
  \BibitemOpen
  \bibfield  {author} {\bibinfo {author} {\bibfnamefont {S.}~\bibnamefont
  {Gopalakrishnan}}\ and\ \bibinfo {author} {\bibfnamefont {B.}~\bibnamefont
  {Zakirov}},\ }\bibfield  {title} {\bibinfo {title} {Facilitated quantum
  cellular automata as simple models with non-thermal eigenstates and
  dynamics},\ }\href {https://doi.org/10.1088/2058-9565/aad759} {\bibfield
  {journal} {\bibinfo  {journal} {Quantum Sci. Technol.}\ }\textbf {\bibinfo
  {volume} {3}},\ \bibinfo {pages} {044004} (\bibinfo {year}
  {2018})}\BibitemShut {NoStop}%
\bibitem [{\citenamefont {Friedman}\ \emph {et~al.}(2019)\citenamefont
  {Friedman}, \citenamefont {Gopalakrishnan},\ and\ \citenamefont
  {Vasseur}}]{Friedman-2019}%
  \BibitemOpen
  \bibfield  {author} {\bibinfo {author} {\bibfnamefont {A.~J.}\ \bibnamefont
  {Friedman}}, \bibinfo {author} {\bibfnamefont {S.}~\bibnamefont
  {Gopalakrishnan}},\ and\ \bibinfo {author} {\bibfnamefont {R.}~\bibnamefont
  {Vasseur}},\ }\bibfield  {title} {\bibinfo {title} {Integrable {{Many-Body
  Quantum Floquet-Thouless Pumps}}},\ }\href
  {https://doi.org/10.1103/PhysRevLett.123.170603} {\bibfield  {journal}
  {\bibinfo  {journal} {Phys. Rev. Lett.}\ }\textbf {\bibinfo {volume} {123}},\
  \bibinfo {pages} {170603} (\bibinfo {year} {2019})}\BibitemShut {NoStop}%
\bibitem [{\citenamefont {Bernien}\ \emph {et~al.}(2017)\citenamefont
  {Bernien}, \citenamefont {Schwartz}, \citenamefont {Keesling}, \citenamefont
  {Levine}, \citenamefont {Omran}, \citenamefont {Pichler}, \citenamefont
  {Choi}, \citenamefont {Zibrov}, \citenamefont {Endres}, \citenamefont
  {Greiner}, \citenamefont {Vuletić},\ and\ \citenamefont
  {Lukin}}]{Bernien-2017}%
  \BibitemOpen
  \bibfield  {author} {\bibinfo {author} {\bibfnamefont {H.}~\bibnamefont
  {Bernien}}, \bibinfo {author} {\bibfnamefont {S.}~\bibnamefont {Schwartz}},
  \bibinfo {author} {\bibfnamefont {A.}~\bibnamefont {Keesling}}, \bibinfo
  {author} {\bibfnamefont {H.}~\bibnamefont {Levine}}, \bibinfo {author}
  {\bibfnamefont {A.}~\bibnamefont {Omran}}, \bibinfo {author} {\bibfnamefont
  {H.}~\bibnamefont {Pichler}}, \bibinfo {author} {\bibfnamefont
  {S.}~\bibnamefont {Choi}}, \bibinfo {author} {\bibfnamefont {A.~S.}\
  \bibnamefont {Zibrov}}, \bibinfo {author} {\bibfnamefont {M.}~\bibnamefont
  {Endres}}, \bibinfo {author} {\bibfnamefont {M.}~\bibnamefont {Greiner}},
  \bibinfo {author} {\bibfnamefont {V.}~\bibnamefont {Vuletić}},\ and\
  \bibinfo {author} {\bibfnamefont {M.~D.}\ \bibnamefont {Lukin}},\ }\bibfield
  {title} {\bibinfo {title} {Probing many-body dynamics on a 51-atom quantum
  simulator},\ }\href {https://doi.org/10.1038/nature24622} {\bibfield
  {journal} {\bibinfo  {journal} {Nature}\ }\textbf {\bibinfo {volume} {551}},\
  \bibinfo {pages} {579} (\bibinfo {year} {2017})}\BibitemShut {NoStop}%
\bibitem [{\citenamefont {Iadecola}\ and\ \citenamefont
  {Vijay}(2020)}]{Iadecola-2020}%
  \BibitemOpen
  \bibfield  {author} {\bibinfo {author} {\bibfnamefont {T.}~\bibnamefont
  {Iadecola}}\ and\ \bibinfo {author} {\bibfnamefont {S.}~\bibnamefont
  {Vijay}},\ }\bibfield  {title} {\bibinfo {title} {Nonergodic quantum dynamics
  from deformations of classical cellular automata},\ }\href
  {https://doi.org/10.1103/PhysRevB.102.180302} {\bibfield  {journal} {\bibinfo
   {journal} {Phys. Rev. B}\ }\textbf {\bibinfo {volume} {102}},\ \bibinfo
  {pages} {180302} (\bibinfo {year} {2020})}\BibitemShut {NoStop}%
\bibitem [{\citenamefont {Mizuta}\ \emph {et~al.}(2020)\citenamefont {Mizuta},
  \citenamefont {Takasan},\ and\ \citenamefont {Kawakami}}]{Mizuta-2020}%
  \BibitemOpen
  \bibfield  {author} {\bibinfo {author} {\bibfnamefont {K.}~\bibnamefont
  {Mizuta}}, \bibinfo {author} {\bibfnamefont {K.}~\bibnamefont {Takasan}},\
  and\ \bibinfo {author} {\bibfnamefont {N.}~\bibnamefont {Kawakami}},\
  }\bibfield  {title} {\bibinfo {title} {Exact {{Floquet}} quantum many-body
  scars under {{Rydberg}} blockade},\ }\href
  {https://doi.org/10.1103/PhysRevResearch.2.033284} {\bibfield  {journal}
  {\bibinfo  {journal} {Phys. Rev. Research}\ }\textbf {\bibinfo {volume}
  {2}},\ \bibinfo {pages} {033284} (\bibinfo {year} {2020})}\BibitemShut
  {NoStop}%
\bibitem [{\citenamefont {Sugiura}\ \emph {et~al.}(2021)\citenamefont
  {Sugiura}, \citenamefont {Kuwahara},\ and\ \citenamefont
  {Saito}}]{Sugiura-2021}%
  \BibitemOpen
  \bibfield  {author} {\bibinfo {author} {\bibfnamefont {S.}~\bibnamefont
  {Sugiura}}, \bibinfo {author} {\bibfnamefont {T.}~\bibnamefont {Kuwahara}},\
  and\ \bibinfo {author} {\bibfnamefont {K.}~\bibnamefont {Saito}},\ }\bibfield
   {title} {\bibinfo {title} {Many-body scar state intrinsic to periodically
  driven system},\ }\href {https://doi.org/10.1103/PhysRevResearch.3.L012010}
  {\bibfield  {journal} {\bibinfo  {journal} {Phys. Rev. Research}\ }\textbf
  {\bibinfo {volume} {3}},\ \bibinfo {pages} {L012010} (\bibinfo {year}
  {2021})}\BibitemShut {NoStop}%
\bibitem [{\citenamefont {Das}(2010)}]{Das-2010}%
  \BibitemOpen
  \bibfield  {author} {\bibinfo {author} {\bibfnamefont {A.}~\bibnamefont
  {Das}},\ }\bibfield  {title} {\bibinfo {title} {Exotic freezing of response
  in a quantum many-body system},\ }\href
  {https://doi.org/10.1103/PhysRevB.82.172402} {\bibfield  {journal} {\bibinfo
  {journal} {Phys. Rev. B}\ }\textbf {\bibinfo {volume} {82}},\ \bibinfo
  {pages} {172402} (\bibinfo {year} {2010})}\BibitemShut {NoStop}%
\bibitem [{\citenamefont {Haldar}\ and\ \citenamefont
  {Das}(2022)}]{Haldar-2022}%
  \BibitemOpen
  \bibfield  {author} {\bibinfo {author} {\bibfnamefont {A.}~\bibnamefont
  {Haldar}}\ and\ \bibinfo {author} {\bibfnamefont {A.}~\bibnamefont {Das}},\
  }\bibfield  {title} {\bibinfo {title} {Statistical mechanics of {{Floquet}}
  quantum matter: Exact and emergent conservation laws},\ }\href
  {https://doi.org/10.1088/1361-648X/ac03d2} {\bibfield  {journal} {\bibinfo
  {journal} {J. Phys.: Condens. Matter}\ }\textbf {\bibinfo {volume} {34}},\
  \bibinfo {pages} {234001} (\bibinfo {year} {2022})}\BibitemShut {NoStop}%
\bibitem [{\citenamefont {Oka}\ and\ \citenamefont
  {Kitamura}(2019)}]{Oka-2019}%
  \BibitemOpen
  \bibfield  {author} {\bibinfo {author} {\bibfnamefont {T.}~\bibnamefont
  {Oka}}\ and\ \bibinfo {author} {\bibfnamefont {S.}~\bibnamefont {Kitamura}},\
  }\bibfield  {title} {\bibinfo {title} {Floquet {{Engineering}} of {{Quantum
  Materials}}},\ }\href
  {https://doi.org/10.1146/annurev-conmatphys-031218-013423} {\bibfield
  {journal} {\bibinfo  {journal} {Annu. Rev. Condens. Matter Phys.}\ }\textbf
  {\bibinfo {volume} {10}},\ \bibinfo {pages} {387} (\bibinfo {year}
  {2019})}\BibitemShut {NoStop}%
\bibitem [{\citenamefont {Harper}\ \emph {et~al.}(2020)\citenamefont {Harper},
  \citenamefont {Roy}, \citenamefont {Rudner},\ and\ \citenamefont
  {Sondhi}}]{Harper-2020}%
  \BibitemOpen
  \bibfield  {author} {\bibinfo {author} {\bibfnamefont {F.}~\bibnamefont
  {Harper}}, \bibinfo {author} {\bibfnamefont {R.}~\bibnamefont {Roy}},
  \bibinfo {author} {\bibfnamefont {M.~S.}\ \bibnamefont {Rudner}},\ and\
  \bibinfo {author} {\bibfnamefont {S.}~\bibnamefont {Sondhi}},\ }\bibfield
  {title} {\bibinfo {title} {Topology and {{Broken Symmetry}} in {{Floquet
  Systems}}},\ }\href
  {https://doi.org/10.1146/annurev-conmatphys-031218-013721} {\bibfield
  {journal} {\bibinfo  {journal} {Annu. Rev. Condens. Matter Phys.}\ }\textbf
  {\bibinfo {volume} {11}},\ \bibinfo {pages} {345} (\bibinfo {year}
  {2020})}\BibitemShut {NoStop}%
\bibitem [{\citenamefont {Yates}\ \emph {et~al.}(2022)\citenamefont {Yates},
  \citenamefont {Abanov},\ and\ \citenamefont {Mitra}}]{Yates-2022}%
  \BibitemOpen
  \bibfield  {author} {\bibinfo {author} {\bibfnamefont {D.~J.}\ \bibnamefont
  {Yates}}, \bibinfo {author} {\bibfnamefont {A.~G.}\ \bibnamefont {Abanov}},\
  and\ \bibinfo {author} {\bibfnamefont {A.}~\bibnamefont {Mitra}},\ }\bibfield
   {title} {\bibinfo {title} {Long-lived period-doubled edge modes of
  interacting and disorder-free {{Floquet}} spin chains},\ }\href
  {https://doi.org/10.1038/s42005-022-00818-1} {\bibfield  {journal} {\bibinfo
  {journal} {Commun Phys}\ }\textbf {\bibinfo {volume} {5}},\ \bibinfo {pages}
  {43} (\bibinfo {year} {2022})}\BibitemShut {NoStop}%
\bibitem [{\citenamefont {Wybo}\ \emph {et~al.}(2021)\citenamefont {Wybo},
  \citenamefont {Pollmann}, \citenamefont {Sondhi},\ and\ \citenamefont
  {You}}]{Wybo-2021}%
  \BibitemOpen
  \bibfield  {author} {\bibinfo {author} {\bibfnamefont {E.}~\bibnamefont
  {Wybo}}, \bibinfo {author} {\bibfnamefont {F.}~\bibnamefont {Pollmann}},
  \bibinfo {author} {\bibfnamefont {S.~L.}\ \bibnamefont {Sondhi}},\ and\
  \bibinfo {author} {\bibfnamefont {Y.}~\bibnamefont {You}},\ }\bibfield
  {title} {\bibinfo {title} {Visualizing quasiparticles from quantum
  entanglement for general one-dimensional phases},\ }\href
  {https://doi.org/10.1103/PhysRevB.103.115120} {\bibfield  {journal} {\bibinfo
   {journal} {Phys. Rev. B}\ }\textbf {\bibinfo {volume} {103}},\ \bibinfo
  {pages} {115120} (\bibinfo {year} {2021})}\BibitemShut {NoStop}%
\bibitem [{\citenamefont {Zeng}\ \emph {et~al.}(2019)\citenamefont {Zeng},
  \citenamefont {Chen}, \citenamefont {Zhou},\ and\ \citenamefont
  {Wen}}]{Zeng-2019}%
  \BibitemOpen
  \bibfield  {author} {\bibinfo {author} {\bibfnamefont {B.}~\bibnamefont
  {Zeng}}, \bibinfo {author} {\bibfnamefont {X.}~\bibnamefont {Chen}}, \bibinfo
  {author} {\bibfnamefont {D.-L.}\ \bibnamefont {Zhou}},\ and\ \bibinfo
  {author} {\bibfnamefont {X.-G.}\ \bibnamefont {Wen}},\ }\href
  {https://doi.org/10.1007/978-1-4939-9084-9} {\emph {\bibinfo {title} {Quantum
  {{Information Meets Quantum Matter}}}}}\ (\bibinfo  {publisher}
  {{Springer}},\ \bibinfo {address} {{New York, NY}},\ \bibinfo {year} {2019})\
  \Eprint {https://arxiv.org/abs/1508.02595} {arXiv:1508.02595} \BibitemShut
  {NoStop}%
\bibitem [{\citenamefont {Nielsen}(2006)}]{Nielsen-2006fv}%
  \BibitemOpen
  \bibfield  {author} {\bibinfo {author} {\bibfnamefont {M.~A.}\ \bibnamefont
  {Nielsen}},\ }\bibfield  {title} {\bibinfo {title} {Cluster-state quantum
  computation},\ }\href
  {https://doi.org/http://dx.doi.org/10.1016/S0034-4877(06)80014-5} {\bibfield
  {journal} {\bibinfo  {journal} {Rep. Math. Phys.}\ }\textbf {\bibinfo
  {volume} {57}},\ \bibinfo {pages} {147} (\bibinfo {year} {2006})}\BibitemShut
  {NoStop}%
\bibitem [{\citenamefont {Raussendorf}\ and\ \citenamefont
  {Wei}(2012)}]{Raussendorf-2012xr}%
  \BibitemOpen
  \bibfield  {author} {\bibinfo {author} {\bibfnamefont {R.}~\bibnamefont
  {Raussendorf}}\ and\ \bibinfo {author} {\bibfnamefont {T.-C.}\ \bibnamefont
  {Wei}},\ }\bibfield  {title} {\bibinfo {title} {Quantum {{Computation}} by
  {{Local Measurement}}},\ }\href
  {https://doi.org/10.1146/annurev-conmatphys-020911-125041} {\bibfield
  {journal} {\bibinfo  {journal} {Annu. Rev. Condens. Matter Phys.}\ }\textbf
  {\bibinfo {volume} {3}},\ \bibinfo {pages} {239} (\bibinfo {year}
  {2012})}\BibitemShut {NoStop}%
\bibitem [{\citenamefont {Haldar}\ \emph {et~al.}(2018)\citenamefont {Haldar},
  \citenamefont {Moessner},\ and\ \citenamefont {Das}}]{Haldar-2018}%
  \BibitemOpen
  \bibfield  {author} {\bibinfo {author} {\bibfnamefont {A.}~\bibnamefont
  {Haldar}}, \bibinfo {author} {\bibfnamefont {R.}~\bibnamefont {Moessner}},\
  and\ \bibinfo {author} {\bibfnamefont {A.}~\bibnamefont {Das}},\ }\bibfield
  {title} {\bibinfo {title} {Onset of {{Floquet}} thermalization},\ }\href
  {https://doi.org/10.1103/PhysRevB.97.245122} {\bibfield  {journal} {\bibinfo
  {journal} {Phys. Rev. B}\ }\textbf {\bibinfo {volume} {97}},\ \bibinfo
  {pages} {245122} (\bibinfo {year} {2018})}\BibitemShut {NoStop}%
\bibitem [{\citenamefont {Haldar}\ \emph {et~al.}(2021)\citenamefont {Haldar},
  \citenamefont {Sen}, \citenamefont {Moessner},\ and\ \citenamefont
  {Das}}]{Haldar-2021}%
  \BibitemOpen
  \bibfield  {author} {\bibinfo {author} {\bibfnamefont {A.}~\bibnamefont
  {Haldar}}, \bibinfo {author} {\bibfnamefont {D.}~\bibnamefont {Sen}},
  \bibinfo {author} {\bibfnamefont {R.}~\bibnamefont {Moessner}},\ and\
  \bibinfo {author} {\bibfnamefont {A.}~\bibnamefont {Das}},\ }\bibfield
  {title} {\bibinfo {title} {Dynamical {{Freezing}} and {{Scar Points}} in
  {{Strongly Driven Floquet Matter}}: {{Resonance}} vs {{Emergent Conservation
  Laws}}},\ }\href {https://doi.org/10.1103/PhysRevX.11.021008} {\bibfield
  {journal} {\bibinfo  {journal} {Phys. Rev. X}\ }\textbf {\bibinfo {volume}
  {11}},\ \bibinfo {pages} {021008} (\bibinfo {year} {2021})}\BibitemShut
  {NoStop}%
\bibitem [{\citenamefont {Sellapillay}\ \emph
  {et~al.}(2022{\natexlab{a}})\citenamefont {Sellapillay}, \citenamefont
  {Verga},\ and\ \citenamefont {Di~Molfetta}}]{Sellapillay-2022b}%
  \BibitemOpen
  \bibfield  {author} {\bibinfo {author} {\bibfnamefont {K.}~\bibnamefont
  {Sellapillay}}, \bibinfo {author} {\bibfnamefont {A.~D.}\ \bibnamefont
  {Verga}},\ and\ \bibinfo {author} {\bibfnamefont {G.}~\bibnamefont
  {Di~Molfetta}},\ }\bibfield  {title} {\bibinfo {title} {Entanglement dynamics
  and ergodicity breaking in a quantum cellular automaton},\ }\href
  {https://doi.org/10.1103/PhysRevB.106.104309} {\bibfield  {journal} {\bibinfo
   {journal} {Phys. Rev. B}\ }\textbf {\bibinfo {volume} {106}},\ \bibinfo
  {pages} {104309} (\bibinfo {year} {2022}{\natexlab{a}})}\BibitemShut
  {NoStop}%
\bibitem [{\citenamefont {Grossmann}\ \emph {et~al.}(1991)\citenamefont
  {Grossmann}, \citenamefont {Dittrich}, \citenamefont {Jung},\ and\
  \citenamefont {Hänggi}}]{Grossmann-1991}%
  \BibitemOpen
  \bibfield  {author} {\bibinfo {author} {\bibfnamefont {F.}~\bibnamefont
  {Grossmann}}, \bibinfo {author} {\bibfnamefont {T.}~\bibnamefont {Dittrich}},
  \bibinfo {author} {\bibfnamefont {P.}~\bibnamefont {Jung}},\ and\ \bibinfo
  {author} {\bibfnamefont {P.}~\bibnamefont {Hänggi}},\ }\bibfield  {title}
  {\bibinfo {title} {Coherent destruction of tunneling},\ }\href
  {https://doi.org/10.1103/PhysRevLett.67.516} {\bibfield  {journal} {\bibinfo
  {journal} {Phys. Rev. Lett.}\ }\textbf {\bibinfo {volume} {67}},\ \bibinfo
  {pages} {516} (\bibinfo {year} {1991})}\BibitemShut {NoStop}%
\bibitem [{\citenamefont {Son}\ \emph {et~al.}(2011)\citenamefont {Son},
  \citenamefont {Amico}, \citenamefont {Fazio}, \citenamefont {Hamma},
  \citenamefont {Pascazio},\ and\ \citenamefont {Vedral}}]{Son-2011}%
  \BibitemOpen
  \bibfield  {author} {\bibinfo {author} {\bibfnamefont {W.}~\bibnamefont
  {Son}}, \bibinfo {author} {\bibfnamefont {L.}~\bibnamefont {Amico}}, \bibinfo
  {author} {\bibfnamefont {R.}~\bibnamefont {Fazio}}, \bibinfo {author}
  {\bibfnamefont {A.}~\bibnamefont {Hamma}}, \bibinfo {author} {\bibfnamefont
  {S.}~\bibnamefont {Pascazio}},\ and\ \bibinfo {author} {\bibfnamefont
  {V.}~\bibnamefont {Vedral}},\ }\bibfield  {title} {\bibinfo {title} {Quantum
  phase transition between cluster and antiferromagnetic states},\ }\href
  {https://doi.org/10.1209/0295-5075/95/50001} {\bibfield  {journal} {\bibinfo
  {journal} {EPL}\ }\textbf {\bibinfo {volume} {95}},\ \bibinfo {pages} {50001}
  (\bibinfo {year} {2011})}\BibitemShut {NoStop}%
\bibitem [{\citenamefont {Kitagawa}(2012)}]{Kitagawa-2012fk}%
  \BibitemOpen
  \bibfield  {author} {\bibinfo {author} {\bibfnamefont {T.}~\bibnamefont
  {Kitagawa}},\ }\bibfield  {title} {\bibinfo {title} {Topological phenomena in
  quantum walks: Elementary introduction to the physics of topological
  phases},\ }\href {https://doi.org/10.1007/s11128-012-0425-4} {\bibfield
  {journal} {\bibinfo  {journal} {Quantum Inf. Process.}\ }\textbf {\bibinfo
  {volume} {11}},\ \bibinfo {pages} {1107} (\bibinfo {year}
  {2012})}\BibitemShut {NoStop}%
\bibitem [{\citenamefont {Verga}(2019)}]{Verga-2019}%
  \BibitemOpen
  \bibfield  {author} {\bibinfo {author} {\bibfnamefont {A.~D.}\ \bibnamefont
  {Verga}},\ }\bibfield  {title} {\bibinfo {title} {Interacting quantum walk on
  a graph},\ }\href {https://doi.org/10.1103/PhysRevE.99.012127} {\bibfield
  {journal} {\bibinfo  {journal} {Phys. Rev. E}\ }\textbf {\bibinfo {volume}
  {99}},\ \bibinfo {pages} {012127} (\bibinfo {year} {2019})}\BibitemShut
  {NoStop}%
\bibitem [{\citenamefont {Verga}\ and\ \citenamefont
  {Elías}(2019)}]{Verga-2019b}%
  \BibitemOpen
  \bibfield  {author} {\bibinfo {author} {\bibfnamefont {A.~D.}\ \bibnamefont
  {Verga}}\ and\ \bibinfo {author} {\bibfnamefont {R.~G.}\ \bibnamefont
  {Elías}},\ }\bibfield  {title} {\bibinfo {title} {Thermal state entanglement
  entropy on a quantum graph},\ }\href
  {https://doi.org/10.1103/PhysRevE.100.062137} {\bibfield  {journal} {\bibinfo
   {journal} {Phys. Rev. E}\ }\textbf {\bibinfo {volume} {100}},\ \bibinfo
  {pages} {062137} (\bibinfo {year} {2019})}\BibitemShut {NoStop}%
\bibitem [{\citenamefont {Heyl}(2015)}]{Heyl-2015}%
  \BibitemOpen
  \bibfield  {author} {\bibinfo {author} {\bibfnamefont {M.}~\bibnamefont
  {Heyl}},\ }\bibfield  {title} {\bibinfo {title} {Scaling and {{Universality}}
  at {{Dynamical Quantum Phase Transitions}}},\ }\href
  {https://doi.org/10.1103/PhysRevLett.115.140602} {\bibfield  {journal}
  {\bibinfo  {journal} {Phys. Rev. Lett.}\ }\textbf {\bibinfo {volume} {115}},\
  \bibinfo {pages} {140602} (\bibinfo {year} {2015})}\BibitemShut {NoStop}%
\bibitem [{\citenamefont {Heyl}\ \emph {et~al.}(2013)\citenamefont {Heyl},
  \citenamefont {Polkovnikov},\ and\ \citenamefont {Kehrein}}]{Heyl-2013}%
  \BibitemOpen
  \bibfield  {author} {\bibinfo {author} {\bibfnamefont {M.}~\bibnamefont
  {Heyl}}, \bibinfo {author} {\bibfnamefont {A.}~\bibnamefont {Polkovnikov}},\
  and\ \bibinfo {author} {\bibfnamefont {S.}~\bibnamefont {Kehrein}},\
  }\bibfield  {title} {\bibinfo {title} {Dynamical {{Quantum Phase
  Transitions}} in the {{Transverse-Field Ising Model}}},\ }\href
  {https://doi.org/10.1103/PhysRevLett.110.135704} {\bibfield  {journal}
  {\bibinfo  {journal} {Phys. Rev. Lett.}\ }\textbf {\bibinfo {volume} {110}},\
  \bibinfo {pages} {135704} (\bibinfo {year} {2013})}\BibitemShut {NoStop}%
\bibitem [{\citenamefont {Heyl}(2018)}]{Heyl-2018}%
  \BibitemOpen
  \bibfield  {author} {\bibinfo {author} {\bibfnamefont {M.}~\bibnamefont
  {Heyl}},\ }\bibfield  {title} {\bibinfo {title} {Dynamical quantum phase
  transitions: A review},\ }\href {https://doi.org/10.1088/1361-6633/aaaf9a}
  {\bibfield  {journal} {\bibinfo  {journal} {Rep. Prog. Phys.}\ }\textbf
  {\bibinfo {volume} {81}},\ \bibinfo {pages} {054001} (\bibinfo {year}
  {2018})}\BibitemShut {NoStop}%
\bibitem [{\citenamefont {De~Nicola}\ \emph {et~al.}(2021)\citenamefont
  {De~Nicola}, \citenamefont {Michailidis},\ and\ \citenamefont
  {Serbyn}}]{DeNicola-2021}%
  \BibitemOpen
  \bibfield  {author} {\bibinfo {author} {\bibfnamefont {S.}~\bibnamefont
  {De~Nicola}}, \bibinfo {author} {\bibfnamefont {A.~A.}\ \bibnamefont
  {Michailidis}},\ and\ \bibinfo {author} {\bibfnamefont {M.}~\bibnamefont
  {Serbyn}},\ }\bibfield  {title} {\bibinfo {title} {Entanglement {{View}} of
  {{Dynamical Quantum Phase Transitions}}},\ }\href
  {https://doi.org/10.1103/PhysRevLett.126.040602} {\bibfield  {journal}
  {\bibinfo  {journal} {Phys. Rev. Lett.}\ }\textbf {\bibinfo {volume} {126}},\
  \bibinfo {pages} {040602} (\bibinfo {year} {2021})}\BibitemShut {NoStop}%
\bibitem [{\citenamefont {Jafari}\ and\ \citenamefont
  {Akbari}(2021)}]{Jafari-2021}%
  \BibitemOpen
  \bibfield  {author} {\bibinfo {author} {\bibfnamefont {R.}~\bibnamefont
  {Jafari}}\ and\ \bibinfo {author} {\bibfnamefont {A.}~\bibnamefont
  {Akbari}},\ }\bibfield  {title} {\bibinfo {title} {Floquet dynamical phase
  transition and entanglement spectrum},\ }\href
  {https://doi.org/10.1103/PhysRevA.103.012204} {\bibfield  {journal} {\bibinfo
   {journal} {Phys. Rev. A}\ }\textbf {\bibinfo {volume} {103}},\ \bibinfo
  {pages} {012204} (\bibinfo {year} {2021})}\BibitemShut {NoStop}%
\bibitem [{\citenamefont {Sellapillay}\ \emph
  {et~al.}(2022{\natexlab{b}})\citenamefont {Sellapillay}, \citenamefont
  {Arrighi},\ and\ \citenamefont {Di~Molfetta}}]{Sellapillay-2022}%
  \BibitemOpen
  \bibfield  {author} {\bibinfo {author} {\bibfnamefont {K.}~\bibnamefont
  {Sellapillay}}, \bibinfo {author} {\bibfnamefont {P.}~\bibnamefont
  {Arrighi}},\ and\ \bibinfo {author} {\bibfnamefont {G.}~\bibnamefont
  {Di~Molfetta}},\ }\bibfield  {title} {\bibinfo {title} {A discrete
  relativistic spacetime formalism for 1 + 1-{{QED}} with continuum limits},\
  }\href {https://doi.org/10.1038/s41598-022-06241-4} {\bibfield  {journal}
  {\bibinfo  {journal} {Sci Rep}\ }\textbf {\bibinfo {volume} {12}},\ \bibinfo
  {pages} {2198} (\bibinfo {year} {2022}{\natexlab{b}})}\BibitemShut {NoStop}%
\bibitem [{\citenamefont {Raussendorf}\ \emph {et~al.}(2005)\citenamefont
  {Raussendorf}, \citenamefont {Bravyi},\ and\ \citenamefont
  {Harrington}}]{Raussendorf-2005}%
  \BibitemOpen
  \bibfield  {author} {\bibinfo {author} {\bibfnamefont {R.}~\bibnamefont
  {Raussendorf}}, \bibinfo {author} {\bibfnamefont {S.}~\bibnamefont
  {Bravyi}},\ and\ \bibinfo {author} {\bibfnamefont {J.}~\bibnamefont
  {Harrington}},\ }\bibfield  {title} {\bibinfo {title} {Long-range quantum
  entanglement in noisy cluster states},\ }\href
  {https://doi.org/10.1103/PhysRevA.71.062313} {\bibfield  {journal} {\bibinfo
  {journal} {Phys. Rev. A}\ }\textbf {\bibinfo {volume} {71}},\ \bibinfo
  {pages} {062313} (\bibinfo {year} {2005})}\BibitemShut {NoStop}%
\bibitem [{\citenamefont {Hein}\ \emph {et~al.}(2006)\citenamefont {Hein},
  \citenamefont {Dür}, \citenamefont {Eisert}, \citenamefont {Raussendorf},
  \citenamefont {Nest},\ and\ \citenamefont {Briegel}}]{Hein-2006eu}%
  \BibitemOpen
  \bibfield  {author} {\bibinfo {author} {\bibfnamefont {M.}~\bibnamefont
  {Hein}}, \bibinfo {author} {\bibfnamefont {W.}~\bibnamefont {Dür}}, \bibinfo
  {author} {\bibfnamefont {J.}~\bibnamefont {Eisert}}, \bibinfo {author}
  {\bibfnamefont {R.}~\bibnamefont {Raussendorf}}, \bibinfo {author}
  {\bibfnamefont {M.}~\bibnamefont {Nest}},\ and\ \bibinfo {author}
  {\bibfnamefont {H.-J.}\ \bibnamefont {Briegel}},\ }\bibfield  {title}
  {\bibinfo {title} {Entanglement in graph states and its applications},\
  }\bibfield  {journal} {\bibinfo  {journal} {ArXiv Prepr.}\ }\href
  {https://doi.org/10.48550/arXiv.quant-ph/0602096}
  {10.48550/arXiv.quant-ph/0602096} (\bibinfo {year} {2006})\BibitemShut
  {NoStop}%
\bibitem [{\citenamefont {Vidal}(2003)}]{Vidal-2003}%
  \BibitemOpen
  \bibfield  {author} {\bibinfo {author} {\bibfnamefont {G.}~\bibnamefont
  {Vidal}},\ }\bibfield  {title} {\bibinfo {title} {Efficient {{Classical
  Simulation}} of {{Slightly Entangled Quantum Computations}}},\ }\href
  {https://doi.org/10.1103/PhysRevLett.91.147902} {\bibfield  {journal}
  {\bibinfo  {journal} {Phys. Rev. Lett.}\ }\textbf {\bibinfo {volume} {91}},\
  \bibinfo {pages} {147902} (\bibinfo {year} {2003})}\BibitemShut {NoStop}%
\bibitem [{\citenamefont {Verresen}\ \emph {et~al.}(2017)\citenamefont
  {Verresen}, \citenamefont {Moessner},\ and\ \citenamefont
  {Pollmann}}]{Verresen-2017}%
  \BibitemOpen
  \bibfield  {author} {\bibinfo {author} {\bibfnamefont {R.}~\bibnamefont
  {Verresen}}, \bibinfo {author} {\bibfnamefont {R.}~\bibnamefont {Moessner}},\
  and\ \bibinfo {author} {\bibfnamefont {F.}~\bibnamefont {Pollmann}},\
  }\bibfield  {title} {\bibinfo {title} {One-dimensional symmetry protected
  topological phases and their transitions},\ }\href
  {https://doi.org/10.1103/PhysRevB.96.165124} {\bibfield  {journal} {\bibinfo
  {journal} {Phys. Rev. B}\ }\textbf {\bibinfo {volume} {96}},\ \bibinfo
  {pages} {165124} (\bibinfo {year} {2017})}\BibitemShut {NoStop}%
\bibitem [{\citenamefont {Suzuki}(1971)}]{Suzuki-1971}%
  \BibitemOpen
  \bibfield  {author} {\bibinfo {author} {\bibfnamefont {M.}~\bibnamefont
  {Suzuki}},\ }\bibfield  {title} {\bibinfo {title} {Relationship among
  {{Exactly Soluble Models}} of {{Critical Phenomena}}. {{I}}: {{2D Ising
  Model}}, {{Dimer Problem}} and the {{Generalized XY-Model}}},\ }\href
  {https://doi.org/10.1143/PTP.46.1337} {\bibfield  {journal} {\bibinfo
  {journal} {Prog. Theor. Phys.}\ }\textbf {\bibinfo {volume} {46}},\ \bibinfo
  {pages} {1337} (\bibinfo {year} {1971})}\BibitemShut {NoStop}%
\bibitem [{\citenamefont {Pachos}\ and\ \citenamefont
  {Plenio}(2004)}]{Pachos-2004}%
  \BibitemOpen
  \bibfield  {author} {\bibinfo {author} {\bibfnamefont {J.~K.}\ \bibnamefont
  {Pachos}}\ and\ \bibinfo {author} {\bibfnamefont {M.~B.}\ \bibnamefont
  {Plenio}},\ }\bibfield  {title} {\bibinfo {title} {Three-{{Spin
  Interactions}} in {{Optical Lattices}} and {{Criticality}} in {{Cluster
  Hamiltonians}}},\ }\href {https://doi.org/10.1103/PhysRevLett.93.056402}
  {\bibfield  {journal} {\bibinfo  {journal} {Phys. Rev. Lett.}\ }\textbf
  {\bibinfo {volume} {93}},\ \bibinfo {pages} {056402} (\bibinfo {year}
  {2004})}\BibitemShut {NoStop}%
\bibitem [{\citenamefont {Skrøvseth}\ and\ \citenamefont
  {Bartlett}(2009)}]{Skrovseth-2009}%
  \BibitemOpen
  \bibfield  {author} {\bibinfo {author} {\bibfnamefont {S.~O.}\ \bibnamefont
  {Skrøvseth}}\ and\ \bibinfo {author} {\bibfnamefont {S.~D.}\ \bibnamefont
  {Bartlett}},\ }\bibfield  {title} {\bibinfo {title} {Phase transitions and
  localizable entanglement in cluster-state spin chains with {{Ising}}
  couplings and local fields},\ }\href
  {https://doi.org/10.1103/PhysRevA.80.022316} {\bibfield  {journal} {\bibinfo
  {journal} {Phys. Rev. A}\ }\textbf {\bibinfo {volume} {80}},\ \bibinfo
  {pages} {022316} (\bibinfo {year} {2009})}\BibitemShut {NoStop}%
\bibitem [{\citenamefont {Smacchia}\ \emph {et~al.}(2011)\citenamefont
  {Smacchia}, \citenamefont {Amico}, \citenamefont {Facchi}, \citenamefont
  {Fazio}, \citenamefont {Florio}, \citenamefont {Pascazio},\ and\
  \citenamefont {Vedral}}]{Smacchia-2011}%
  \BibitemOpen
  \bibfield  {author} {\bibinfo {author} {\bibfnamefont {P.}~\bibnamefont
  {Smacchia}}, \bibinfo {author} {\bibfnamefont {L.}~\bibnamefont {Amico}},
  \bibinfo {author} {\bibfnamefont {P.}~\bibnamefont {Facchi}}, \bibinfo
  {author} {\bibfnamefont {R.}~\bibnamefont {Fazio}}, \bibinfo {author}
  {\bibfnamefont {G.}~\bibnamefont {Florio}}, \bibinfo {author} {\bibfnamefont
  {S.}~\bibnamefont {Pascazio}},\ and\ \bibinfo {author} {\bibfnamefont
  {V.}~\bibnamefont {Vedral}},\ }\bibfield  {title} {\bibinfo {title}
  {Statistical mechanics of the cluster {{Ising}} model},\ }\href
  {https://doi.org/10.1103/PhysRevA.84.022304} {\bibfield  {journal} {\bibinfo
  {journal} {Phys. Rev. A}\ }\textbf {\bibinfo {volume} {84}},\ \bibinfo
  {pages} {022304} (\bibinfo {year} {2011})}\BibitemShut {NoStop}%
\bibitem [{\citenamefont {Montes}\ and\ \citenamefont
  {Hamma}(2012)}]{Montes-2012}%
  \BibitemOpen
  \bibfield  {author} {\bibinfo {author} {\bibfnamefont {S.}~\bibnamefont
  {Montes}}\ and\ \bibinfo {author} {\bibfnamefont {A.}~\bibnamefont {Hamma}},\
  }\bibfield  {title} {\bibinfo {title} {Phase diagram and quench dynamics of
  the cluster-{{XY}} spin chain},\ }\href
  {https://doi.org/10.1103/PhysRevE.86.021101} {\bibfield  {journal} {\bibinfo
  {journal} {Phys. Rev. E}\ }\textbf {\bibinfo {volume} {86}},\ \bibinfo
  {pages} {021101} (\bibinfo {year} {2012})}\BibitemShut {NoStop}%
\bibitem [{\citenamefont {Lakshminarayan}\ and\ \citenamefont
  {Subrahmanyam}(2005)}]{Lakshminarayan-2005}%
  \BibitemOpen
  \bibfield  {author} {\bibinfo {author} {\bibfnamefont {A.}~\bibnamefont
  {Lakshminarayan}}\ and\ \bibinfo {author} {\bibfnamefont {V.}~\bibnamefont
  {Subrahmanyam}},\ }\bibfield  {title} {\bibinfo {title} {Multipartite
  entanglement in a one-dimensional time-dependent {{Ising}} model},\ }\href
  {https://doi.org/10.1103/PhysRevA.71.062334} {\bibfield  {journal} {\bibinfo
  {journal} {Phys. Rev. A}\ }\textbf {\bibinfo {volume} {71}},\ \bibinfo
  {pages} {062334} (\bibinfo {year} {2005})}\BibitemShut {NoStop}%
\bibitem [{\citenamefont {Jordan}\ and\ \citenamefont
  {Wigner}(1928)}]{Jordan-1928}%
  \BibitemOpen
  \bibfield  {author} {\bibinfo {author} {\bibfnamefont {P.}~\bibnamefont
  {Jordan}}\ and\ \bibinfo {author} {\bibfnamefont {E.}~\bibnamefont
  {Wigner}},\ }\bibfield  {title} {\bibinfo {title} {{Über das Paulische
  Äquivalenzverbot}},\ }\href {https://doi.org/10.1007/BF01331938} {\bibfield
  {journal} {\bibinfo  {journal} {Z. Physik}\ }\textbf {\bibinfo {volume}
  {47}},\ \bibinfo {pages} {631} (\bibinfo {year} {1928})}\BibitemShut
  {NoStop}%
\bibitem [{\citenamefont {Mbeng}\ \emph {et~al.}(2020)\citenamefont {Mbeng},
  \citenamefont {Russomanno},\ and\ \citenamefont {Santoro}}]{Mbeng-2020}%
  \BibitemOpen
  \bibfield  {author} {\bibinfo {author} {\bibfnamefont {G.~B.}\ \bibnamefont
  {Mbeng}}, \bibinfo {author} {\bibfnamefont {A.}~\bibnamefont {Russomanno}},\
  and\ \bibinfo {author} {\bibfnamefont {G.~E.}\ \bibnamefont {Santoro}},\
  }\bibfield  {title} {\bibinfo {title} {The quantum {{Ising}} chain for
  beginners},\ }\bibfield  {journal} {\bibinfo  {journal} {arXiv:2009.09208}\
  }\href {https://doi.org/10.48550/arXiv.2009.09208}
  {10.48550/arXiv.2009.09208} (\bibinfo {year} {2020}),\ \Eprint
  {https://arxiv.org/abs/2009.09208} {arXiv:2009.09208} \BibitemShut {NoStop}%
\bibitem [{\citenamefont {Asbóth}\ \emph {et~al.}(2016)\citenamefont
  {Asbóth}, \citenamefont {Oroszlány},\ and\ \citenamefont
  {Pályi}}]{Asboth-2016zr}%
  \BibitemOpen
  \bibfield  {author} {\bibinfo {author} {\bibfnamefont {J.~K.}\ \bibnamefont
  {Asbóth}}, \bibinfo {author} {\bibfnamefont {L.}~\bibnamefont
  {Oroszlány}},\ and\ \bibinfo {author} {\bibfnamefont {A.}~\bibnamefont
  {Pályi}},\ }\href {https://doi.org/10.1007/978-3-319-25607-8} {\emph
  {\bibinfo {title} {A {{Short Course}} on {{Topological Insulators}}}}},\
  \bibinfo {series} {Lecture {{Notes}} in {{Physics}}}, Vol.\ \bibinfo {volume}
  {919}\ (\bibinfo  {publisher} {{Springer}},\ \bibinfo {year}
  {2016})\BibitemShut {NoStop}%
\bibitem [{\citenamefont {Roy}\ and\ \citenamefont {Harper}(2017)}]{Roy-2017a}%
  \BibitemOpen
  \bibfield  {author} {\bibinfo {author} {\bibfnamefont {R.}~\bibnamefont
  {Roy}}\ and\ \bibinfo {author} {\bibfnamefont {F.}~\bibnamefont {Harper}},\
  }\bibfield  {title} {\bibinfo {title} {Periodic table for {{Floquet}}
  topological insulators},\ }\href {https://doi.org/10.1103/PhysRevB.96.155118}
  {\bibfield  {journal} {\bibinfo  {journal} {Phys. Rev. B}\ }\textbf {\bibinfo
  {volume} {96}},\ \bibinfo {pages} {155118} (\bibinfo {year}
  {2017})}\BibitemShut {NoStop}%
\bibitem [{\citenamefont {Asbóth}\ and\ \citenamefont
  {Obuse}(2013)}]{Asboth-2013yq}%
  \BibitemOpen
  \bibfield  {author} {\bibinfo {author} {\bibfnamefont {J.~K.}\ \bibnamefont
  {Asbóth}}\ and\ \bibinfo {author} {\bibfnamefont {H.}~\bibnamefont
  {Obuse}},\ }\bibfield  {title} {\bibinfo {title} {Bulk-boundary
  correspondence for chiral symmetric quantum walks},\ }\href
  {https://doi.org/10.1103/PhysRevB.88.121406} {\bibfield  {journal} {\bibinfo
  {journal} {Phys. Rev. B}\ }\textbf {\bibinfo {volume} {88}},\ \bibinfo
  {pages} {121406} (\bibinfo {year} {2013})}\BibitemShut {NoStop}%
\bibitem [{\citenamefont {Zhou}\ and\ \citenamefont {Gong}(2018)}]{Zhou-2018}%
  \BibitemOpen
  \bibfield  {author} {\bibinfo {author} {\bibfnamefont {L.}~\bibnamefont
  {Zhou}}\ and\ \bibinfo {author} {\bibfnamefont {J.}~\bibnamefont {Gong}},\
  }\bibfield  {title} {\bibinfo {title} {Floquet topological phases in a
  spin-1/2 double kicked rotor},\ }\href
  {https://doi.org/10.1103/PhysRevA.97.063603} {\bibfield  {journal} {\bibinfo
  {journal} {Phys. Rev. A}\ }\textbf {\bibinfo {volume} {97}},\ \bibinfo
  {pages} {063603} (\bibinfo {year} {2018})}\BibitemShut {NoStop}%
\bibitem [{\citenamefont {Ohta}\ \emph {et~al.}(2016)\citenamefont {Ohta},
  \citenamefont {Tanaka}, \citenamefont {Danshita},\ and\ \citenamefont
  {Totsuka}}]{Ohta-2016}%
  \BibitemOpen
  \bibfield  {author} {\bibinfo {author} {\bibfnamefont {T.}~\bibnamefont
  {Ohta}}, \bibinfo {author} {\bibfnamefont {S.}~\bibnamefont {Tanaka}},
  \bibinfo {author} {\bibfnamefont {I.}~\bibnamefont {Danshita}},\ and\
  \bibinfo {author} {\bibfnamefont {K.}~\bibnamefont {Totsuka}},\ }\bibfield
  {title} {\bibinfo {title} {Topological and dynamical properties of a
  generalized cluster model in one dimension},\ }\href
  {https://doi.org/10.1103/PhysRevB.93.165423} {\bibfield  {journal} {\bibinfo
  {journal} {Phys. Rev. B}\ }\textbf {\bibinfo {volume} {93}},\ \bibinfo
  {pages} {165423} (\bibinfo {year} {2016})}\BibitemShut {NoStop}%
\bibitem [{\citenamefont {Brennen}(2003)}]{Brennen-2003a}%
  \BibitemOpen
  \bibfield  {author} {\bibinfo {author} {\bibfnamefont {G.~K.}\ \bibnamefont
  {Brennen}},\ }\bibfield  {title} {\bibinfo {title} {An observable measure of
  entanglement for pure states of multi-qubit systems},\ }\href
  {https://doi.org/10.26421/QIC3.6-5} {\bibfield  {journal} {\bibinfo
  {journal} {Quantum Info. Comput.}\ }\textbf {\bibinfo {volume} {3}},\
  \bibinfo {pages} {619} (\bibinfo {year} {2003})}\BibitemShut {NoStop}%
\bibitem [{\citenamefont {Meyer}\ and\ \citenamefont
  {Wallach}(2002)}]{Meyer-2002}%
  \BibitemOpen
  \bibfield  {author} {\bibinfo {author} {\bibfnamefont {D.~A.}\ \bibnamefont
  {Meyer}}\ and\ \bibinfo {author} {\bibfnamefont {N.~R.}\ \bibnamefont
  {Wallach}},\ }\bibfield  {title} {\bibinfo {title} {Global entanglement in
  multiparticle systems},\ }\href {https://doi.org/10.1063/1.1497700}
  {\bibfield  {journal} {\bibinfo  {journal} {J. Math. Phys.}\ }\textbf
  {\bibinfo {volume} {43}},\ \bibinfo {pages} {4273} (\bibinfo {year}
  {2002})}\BibitemShut {NoStop}%
\bibitem [{\citenamefont {Giraud}\ \emph {et~al.}(2007)\citenamefont {Giraud},
  \citenamefont {Martin},\ and\ \citenamefont {Georgeot}}]{Giraud-2007}%
  \BibitemOpen
  \bibfield  {author} {\bibinfo {author} {\bibfnamefont {O.}~\bibnamefont
  {Giraud}}, \bibinfo {author} {\bibfnamefont {J.}~\bibnamefont {Martin}},\
  and\ \bibinfo {author} {\bibfnamefont {B.}~\bibnamefont {Georgeot}},\
  }\bibfield  {title} {\bibinfo {title} {Entanglement of localized states},\
  }\href {https://doi.org/10.1103/PhysRevA.76.042333} {\bibfield  {journal}
  {\bibinfo  {journal} {Phys. Rev. A}\ }\textbf {\bibinfo {volume} {76}},\
  \bibinfo {pages} {042333} (\bibinfo {year} {2007})}\BibitemShut {NoStop}%
\bibitem [{\citenamefont {Radgohar}\ and\ \citenamefont
  {Montakhab}(2018)}]{Radgohar-2018}%
  \BibitemOpen
  \bibfield  {author} {\bibinfo {author} {\bibfnamefont {R.}~\bibnamefont
  {Radgohar}}\ and\ \bibinfo {author} {\bibfnamefont {A.}~\bibnamefont
  {Montakhab}},\ }\bibfield  {title} {\bibinfo {title} {Global entanglement and
  quantum phase transitions in the transverse {{XY Heisenberg}} chain},\ }\href
  {https://doi.org/10.1103/PhysRevB.97.024434} {\bibfield  {journal} {\bibinfo
  {journal} {Phys. Rev. B}\ }\textbf {\bibinfo {volume} {97}},\ \bibinfo
  {pages} {024434} (\bibinfo {year} {2018})}\BibitemShut {NoStop}%
\bibitem [{\citenamefont {Häppölä}\ \emph {et~al.}(2012)\citenamefont
  {Häppölä}, \citenamefont {Halász},\ and\ \citenamefont
  {Hamma}}]{Happola-2012}%
  \BibitemOpen
  \bibfield  {author} {\bibinfo {author} {\bibfnamefont {J.}~\bibnamefont
  {Häppölä}}, \bibinfo {author} {\bibfnamefont {G.~B.}\ \bibnamefont
  {Halász}},\ and\ \bibinfo {author} {\bibfnamefont {A.}~\bibnamefont
  {Hamma}},\ }\bibfield  {title} {\bibinfo {title} {Universality and robustness
  of revivals in the transverse field {{XY}} model},\ }\href
  {https://doi.org/10.1103/PhysRevA.85.032114} {\bibfield  {journal} {\bibinfo
  {journal} {Phys. Rev. A}\ }\textbf {\bibinfo {volume} {85}},\ \bibinfo
  {pages} {032114} (\bibinfo {year} {2012})}\BibitemShut {NoStop}%
\bibitem [{\citenamefont {Vanicat}\ \emph {et~al.}(2018)\citenamefont
  {Vanicat}, \citenamefont {Zadnik},\ and\ \citenamefont
  {Prosen}}]{Vanicat-2018}%
  \BibitemOpen
  \bibfield  {author} {\bibinfo {author} {\bibfnamefont {M.}~\bibnamefont
  {Vanicat}}, \bibinfo {author} {\bibfnamefont {L.}~\bibnamefont {Zadnik}},\
  and\ \bibinfo {author} {\bibfnamefont {T.}~\bibnamefont {Prosen}},\
  }\bibfield  {title} {\bibinfo {title} {Integrable {{Trotterization}}: {{Local
  Conservation Laws}} and {{Boundary Driving}}},\ }\href
  {https://doi.org/10.1103/PhysRevLett.121.030606} {\bibfield  {journal}
  {\bibinfo  {journal} {Phys. Rev. Lett.}\ }\textbf {\bibinfo {volume} {121}},\
  \bibinfo {pages} {030606} (\bibinfo {year} {2018})}\BibitemShut {NoStop}%
\bibitem [{\citenamefont {Pai}\ and\ \citenamefont {Pretko}(2019)}]{Pai-2019}%
  \BibitemOpen
  \bibfield  {author} {\bibinfo {author} {\bibfnamefont {S.}~\bibnamefont
  {Pai}}\ and\ \bibinfo {author} {\bibfnamefont {M.}~\bibnamefont {Pretko}},\
  }\bibfield  {title} {\bibinfo {title} {Dynamical {{Scar States}} in {{Driven
  Fracton Systems}}},\ }\href {https://doi.org/10.1103/PhysRevLett.123.136401}
  {\bibfield  {journal} {\bibinfo  {journal} {Phys. Rev. Lett.}\ }\textbf
  {\bibinfo {volume} {123}},\ \bibinfo {pages} {136401} (\bibinfo {year}
  {2019})}\BibitemShut {NoStop}%
\bibitem [{\citenamefont {Sellapillay}\ and\ \citenamefont
  {Verga}(2021)}]{Sellapillay-2021}%
  \BibitemOpen
  \bibfield  {author} {\bibinfo {author} {\bibfnamefont {K.}~\bibnamefont
  {Sellapillay}}\ and\ \bibinfo {author} {\bibfnamefont {A.~D.}\ \bibnamefont
  {Verga}},\ }\bibfield  {title} {\bibinfo {title} {Quantum walk on a graph of
  spins: {{Magnetism}} and entanglement},\ }\href
  {https://doi.org/10.1103/PhysRevE.103.032123} {\bibfield  {journal} {\bibinfo
   {journal} {Phys. Rev. E}\ }\textbf {\bibinfo {volume} {103}},\ \bibinfo
  {pages} {032123} (\bibinfo {year} {2021})}\BibitemShut {NoStop}%
\bibitem [{\citenamefont {Meyer}(1996)}]{Meyer-1996sf}%
  \BibitemOpen
  \bibfield  {author} {\bibinfo {author} {\bibfnamefont {D.~A.}\ \bibnamefont
  {Meyer}},\ }\bibfield  {title} {\bibinfo {title} {From quantum cellular
  automata to quantum lattice gases},\ }\href
  {https://doi.org/10.1007/BF02199356} {\bibfield  {journal} {\bibinfo
  {journal} {J. Stat. Phys.}\ }\textbf {\bibinfo {volume} {85}},\ \bibinfo
  {pages} {551} (\bibinfo {year} {1996})}\BibitemShut {NoStop}%
\bibitem [{\citenamefont {Strauch}(2006)}]{Strauch-2006}%
  \BibitemOpen
  \bibfield  {author} {\bibinfo {author} {\bibfnamefont {F.~W.}\ \bibnamefont
  {Strauch}},\ }\bibfield  {title} {\bibinfo {title} {Relativistic quantum
  walks},\ }\href {https://doi.org/10.1103/PhysRevA.73.054302} {\bibfield
  {journal} {\bibinfo  {journal} {Phys. Rev. A}\ }\textbf {\bibinfo {volume}
  {73}},\ \bibinfo {pages} {054302} (\bibinfo {year} {2006})}\BibitemShut
  {NoStop}%
\bibitem [{\citenamefont {Di~Molfetta}\ and\ \citenamefont
  {Debbasch}(2012)}]{Di-Molfetta-2012fv}%
  \BibitemOpen
  \bibfield  {author} {\bibinfo {author} {\bibfnamefont {G.}~\bibnamefont
  {Di~Molfetta}}\ and\ \bibinfo {author} {\bibfnamefont {F.}~\bibnamefont
  {Debbasch}},\ }\bibfield  {title} {\bibinfo {title} {Discrete-time quantum
  walks: {{Continuous}} limit and symmetries},\ }\href
  {https://doi.org/http://dx.doi.org/10.1063/1.4764876} {\bibfield  {journal}
  {\bibinfo  {journal} {J. Math. Phys.}\ }\textbf {\bibinfo {volume} {53}},\
  \bibinfo {pages} {123302} (\bibinfo {year} {2012})}\BibitemShut {NoStop}%
\bibitem [{\citenamefont {Caux}\ and\ \citenamefont
  {Mossel}(2011)}]{Caux-2011}%
  \BibitemOpen
  \bibfield  {author} {\bibinfo {author} {\bibfnamefont {J.-S.}\ \bibnamefont
  {Caux}}\ and\ \bibinfo {author} {\bibfnamefont {J.}~\bibnamefont {Mossel}},\
  }\bibfield  {title} {\bibinfo {title} {Remarks on the notion of quantum
  integrability},\ }\href {https://doi.org/10.1088/1742-5468/2011/02/P02023}
  {\bibfield  {journal} {\bibinfo  {journal} {J. Stat. Mech.}\ }\textbf
  {\bibinfo {volume} {2011}},\ \bibinfo {pages} {P02023} (\bibinfo {year}
  {2011})}\BibitemShut {NoStop}%
\bibitem [{\citenamefont {Klinovaja}\ \emph {et~al.}(2013)\citenamefont
  {Klinovaja}, \citenamefont {Stano}, \citenamefont {Yazdani},\ and\
  \citenamefont {Loss}}]{Klinovaja-2013}%
  \BibitemOpen
  \bibfield  {author} {\bibinfo {author} {\bibfnamefont {J.}~\bibnamefont
  {Klinovaja}}, \bibinfo {author} {\bibfnamefont {P.}~\bibnamefont {Stano}},
  \bibinfo {author} {\bibfnamefont {A.}~\bibnamefont {Yazdani}},\ and\ \bibinfo
  {author} {\bibfnamefont {D.}~\bibnamefont {Loss}},\ }\bibfield  {title}
  {\bibinfo {title} {Topological {{Superconductivity}} and {{Majorana
  Fermions}} in {{RKKY Systems}}},\ }\href
  {https://doi.org/10.1103/PhysRevLett.111.186805} {\bibfield  {journal}
  {\bibinfo  {journal} {Phys. Rev. Lett.}\ }\textbf {\bibinfo {volume} {111}},\
  \bibinfo {pages} {186805} (\bibinfo {year} {2013})}\BibitemShut {NoStop}%
\bibitem [{\citenamefont {Peres}(1984)}]{Peres-1984ai}%
  \BibitemOpen
  \bibfield  {author} {\bibinfo {author} {\bibfnamefont {A.}~\bibnamefont
  {Peres}},\ }\bibfield  {title} {\bibinfo {title} {Stability of quantum motion
  in chaotic and regular systems},\ }\href
  {https://doi.org/10.1103/PhysRevA.30.1610} {\bibfield  {journal} {\bibinfo
  {journal} {Phys. Rev. A}\ }\textbf {\bibinfo {volume} {30}},\ \bibinfo
  {pages} {1610} (\bibinfo {year} {1984})}\BibitemShut {NoStop}%
\bibitem [{\citenamefont {Heyl}\ and\ \citenamefont
  {Budich}(2017)}]{Heyl-2017}%
  \BibitemOpen
  \bibfield  {author} {\bibinfo {author} {\bibfnamefont {M.}~\bibnamefont
  {Heyl}}\ and\ \bibinfo {author} {\bibfnamefont {J.~C.}\ \bibnamefont
  {Budich}},\ }\bibfield  {title} {\bibinfo {title} {Dynamical topological
  quantum phase transitions for mixed states},\ }\href
  {https://doi.org/10.1103/PhysRevB.96.180304} {\bibfield  {journal} {\bibinfo
  {journal} {Phys. Rev. B}\ }\textbf {\bibinfo {volume} {96}},\ \bibinfo
  {pages} {180304} (\bibinfo {year} {2017})}\BibitemShut {NoStop}%
\bibitem [{\citenamefont {Bhattacharya}\ \emph {et~al.}(2017)\citenamefont
  {Bhattacharya}, \citenamefont {Bandyopadhyay},\ and\ \citenamefont
  {Dutta}}]{Bhattacharya-2017}%
  \BibitemOpen
  \bibfield  {author} {\bibinfo {author} {\bibfnamefont {U.}~\bibnamefont
  {Bhattacharya}}, \bibinfo {author} {\bibfnamefont {S.}~\bibnamefont
  {Bandyopadhyay}},\ and\ \bibinfo {author} {\bibfnamefont {A.}~\bibnamefont
  {Dutta}},\ }\bibfield  {title} {\bibinfo {title} {Mixed state dynamical
  quantum phase transitions},\ }\href
  {https://doi.org/10.1103/PhysRevB.96.180303} {\bibfield  {journal} {\bibinfo
  {journal} {Phys. Rev. B}\ }\textbf {\bibinfo {volume} {96}},\ \bibinfo
  {pages} {180303} (\bibinfo {year} {2017})}\BibitemShut {NoStop}%
\bibitem [{\citenamefont {Coffman}\ \emph {et~al.}(2000)\citenamefont
  {Coffman}, \citenamefont {Kundu},\ and\ \citenamefont
  {Wootters}}]{Coffman-2000}%
  \BibitemOpen
  \bibfield  {author} {\bibinfo {author} {\bibfnamefont {V.}~\bibnamefont
  {Coffman}}, \bibinfo {author} {\bibfnamefont {J.}~\bibnamefont {Kundu}},\
  and\ \bibinfo {author} {\bibfnamefont {W.~K.}\ \bibnamefont {Wootters}},\
  }\bibfield  {title} {\bibinfo {title} {Distributed entanglement},\ }\href
  {https://doi.org/10.1103/PhysRevA.61.052306} {\bibfield  {journal} {\bibinfo
  {journal} {Phys. Rev. A}\ }\textbf {\bibinfo {volume} {61}},\ \bibinfo
  {pages} {052306} (\bibinfo {year} {2000})}\BibitemShut {NoStop}%
\bibitem [{\citenamefont {Asbóth}(2012)}]{Asboth-2012qy}%
  \BibitemOpen
  \bibfield  {author} {\bibinfo {author} {\bibfnamefont {J.~K.}\ \bibnamefont
  {Asbóth}},\ }\bibfield  {title} {\bibinfo {title} {Symmetries, topological
  phases, and bound states in the one-dimensional quantum walk},\ }\href
  {https://doi.org/10.1103/PhysRevB.86.195414} {\bibfield  {journal} {\bibinfo
  {journal} {Phys. Rev. B}\ }\textbf {\bibinfo {volume} {86}},\ \bibinfo
  {pages} {195414} (\bibinfo {year} {2012})}\BibitemShut {NoStop}%
\bibitem [{\citenamefont {Bukov}\ \emph {et~al.}(2015)\citenamefont {Bukov},
  \citenamefont {D'Alessio},\ and\ \citenamefont {Polkovnikov}}]{Bukov-2015}%
  \BibitemOpen
  \bibfield  {author} {\bibinfo {author} {\bibfnamefont {M.}~\bibnamefont
  {Bukov}}, \bibinfo {author} {\bibfnamefont {L.}~\bibnamefont {D'Alessio}},\
  and\ \bibinfo {author} {\bibfnamefont {A.}~\bibnamefont {Polkovnikov}},\
  }\bibfield  {title} {\bibinfo {title} {Universal high-frequency behavior of
  periodically driven systems: From dynamical stabilization to {{Floquet}}
  engineering},\ }\href {https://doi.org/10.1080/00018732.2015.1055918}
  {\bibfield  {journal} {\bibinfo  {journal} {Adv. Phys.}\ }\textbf {\bibinfo
  {volume} {64}},\ \bibinfo {pages} {139} (\bibinfo {year} {2015})}\BibitemShut
  {NoStop}%
\bibitem [{\citenamefont {Hollerith}\ \emph {et~al.}(2022)\citenamefont
  {Hollerith}, \citenamefont {Srakaew}, \citenamefont {Wei}, \citenamefont
  {{Rubio-Abadal}}, \citenamefont {Adler}, \citenamefont {Weckesser},
  \citenamefont {Kruckenhauser}, \citenamefont {Walther}, \citenamefont {{van
  Bijnen}}, \citenamefont {Rui}, \citenamefont {Gross}, \citenamefont {Bloch},\
  and\ \citenamefont {Zeiher}}]{Hollerith-2022}%
  \BibitemOpen
  \bibfield  {author} {\bibinfo {author} {\bibfnamefont {S.}~\bibnamefont
  {Hollerith}}, \bibinfo {author} {\bibfnamefont {K.}~\bibnamefont {Srakaew}},
  \bibinfo {author} {\bibfnamefont {D.}~\bibnamefont {Wei}}, \bibinfo {author}
  {\bibfnamefont {A.}~\bibnamefont {{Rubio-Abadal}}}, \bibinfo {author}
  {\bibfnamefont {D.}~\bibnamefont {Adler}}, \bibinfo {author} {\bibfnamefont
  {P.}~\bibnamefont {Weckesser}}, \bibinfo {author} {\bibfnamefont
  {A.}~\bibnamefont {Kruckenhauser}}, \bibinfo {author} {\bibfnamefont
  {V.}~\bibnamefont {Walther}}, \bibinfo {author} {\bibfnamefont
  {R.}~\bibnamefont {{van Bijnen}}}, \bibinfo {author} {\bibfnamefont
  {J.}~\bibnamefont {Rui}}, \bibinfo {author} {\bibfnamefont {C.}~\bibnamefont
  {Gross}}, \bibinfo {author} {\bibfnamefont {I.}~\bibnamefont {Bloch}},\ and\
  \bibinfo {author} {\bibfnamefont {J.}~\bibnamefont {Zeiher}},\ }\bibfield
  {title} {\bibinfo {title} {Realizing {{Distance-Selective Interactions}} in a
  {{Rydberg-Dressed Atom Array}}},\ }\href
  {https://doi.org/10.1103/PhysRevLett.128.113602} {\bibfield  {journal}
  {\bibinfo  {journal} {Phys. Rev. Lett.}\ }\textbf {\bibinfo {volume} {128}},\
  \bibinfo {pages} {113602} (\bibinfo {year} {2022})}\BibitemShut {NoStop}%
\bibitem [{\citenamefont {Khazali}(2022)}]{Khazali-2022}%
  \BibitemOpen
  \bibfield  {author} {\bibinfo {author} {\bibfnamefont {M.}~\bibnamefont
  {Khazali}},\ }\bibfield  {title} {\bibinfo {title} {Discrete-{{Time
  Quantum-Walk}} \& {{Floquet Topological Insulators}} via {{Distance-Selective
  Rydberg-Interaction}}},\ }\href {https://doi.org/10.22331/q-2022-03-03-664}
  {\bibfield  {journal} {\bibinfo  {journal} {Quantum}\ }\textbf {\bibinfo
  {volume} {6}},\ \bibinfo {pages} {664} (\bibinfo {year} {2022})}\BibitemShut
  {NoStop}%
\bibitem [{\citenamefont {Brun}\ and\ \citenamefont
  {Mlodinow}(2020)}]{Brun-2020}%
  \BibitemOpen
  \bibfield  {author} {\bibinfo {author} {\bibfnamefont {T.~A.}\ \bibnamefont
  {Brun}}\ and\ \bibinfo {author} {\bibfnamefont {L.}~\bibnamefont
  {Mlodinow}},\ }\bibfield  {title} {\bibinfo {title} {Quantum cellular
  automata and quantum field theory in two spatial dimensions},\ }\href
  {https://doi.org/10.1103/PhysRevA.102.062222} {\bibfield  {journal} {\bibinfo
   {journal} {Phys. Rev. A}\ }\textbf {\bibinfo {volume} {102}},\ \bibinfo
  {pages} {062222} (\bibinfo {year} {2020})}\BibitemShut {NoStop}%
\bibitem [{\citenamefont {Lee}\ and\ \citenamefont {Vidal}(2013)}]{Lee-2013}%
  \BibitemOpen
  \bibfield  {author} {\bibinfo {author} {\bibfnamefont {Y.~A.}\ \bibnamefont
  {Lee}}\ and\ \bibinfo {author} {\bibfnamefont {G.}~\bibnamefont {Vidal}},\
  }\bibfield  {title} {\bibinfo {title} {Entanglement negativity and
  topological order},\ }\href {https://doi.org/10.1103/PhysRevA.88.042318}
  {\bibfield  {journal} {\bibinfo  {journal} {Phys. Rev. A}\ }\textbf {\bibinfo
  {volume} {88}},\ \bibinfo {pages} {042318} (\bibinfo {year}
  {2013})}\BibitemShut {NoStop}%
\bibitem [{\citenamefont {Kittel}(2018)}]{Kittel-2018}%
  \BibitemOpen
  \bibfield  {author} {\bibinfo {author} {\bibfnamefont {C.}~\bibnamefont
  {Kittel}},\ }\href@noop {} {\emph {\bibinfo {title} {Introduction to {{Solid
  State Physics}}}}},\ \bibinfo {edition} {9th}\ ed.\ (\bibinfo  {publisher}
  {{Wiley-VCH}},\ \bibinfo {year} {2018})\BibitemShut {NoStop}%
\bibitem [{\citenamefont {Ruderman}\ and\ \citenamefont
  {Kittel}(1954)}]{Ruderman-1954}%
  \BibitemOpen
  \bibfield  {author} {\bibinfo {author} {\bibfnamefont {M.~A.}\ \bibnamefont
  {Ruderman}}\ and\ \bibinfo {author} {\bibfnamefont {C.}~\bibnamefont
  {Kittel}},\ }\bibfield  {title} {\bibinfo {title} {Indirect {{Exchange
  Coupling}} of {{Nuclear Magnetic Moments}} by {{Conduction Electrons}}},\
  }\href {https://doi.org/10.1103/PhysRev.96.99} {\bibfield  {journal}
  {\bibinfo  {journal} {Phys. Rev.}\ }\textbf {\bibinfo {volume} {96}},\
  \bibinfo {pages} {99} (\bibinfo {year} {1954})}\BibitemShut {NoStop}%
\end{thebibliography}
\end{document}